\documentclass[prd,aps,a4paper,twocolumn,nofootinbib,floatfix,eqsecnum]{revtex4}  

\newif\ifusesec
\usesectrue  
   
\usepackage{graphicx} 
\usepackage{amsmath,amsfonts,amssymb}

\newcommand{\beq}{\begin{equation}}
\newcommand{\eeq}{\end{equation}}
\newcommand{\bea}{\begin{eqnarray}}
\newcommand{\eea}{\end{eqnarray}}
\newcommand{\eq}{\eqref}

\newcommand{\Int}{\int\limits}

\newcommand{\g}{{\gamma}}

\newcommand{\pinf}{p_{\infty}}

\def\Q#1{\mathrm{Q}_{#1}}

\def\cat#1{{\beta(#1)}}
\def\catd{{\rm K}}

\begin{document}

\title{
Gravitational scattering at the seventh order in $G$: \\
nonlocal contribution at the sixth post-Newtonian accuracy
}

\author{Donato Bini$^{1,2}$, Thibault Damour$^3$, Andrea Geralico$^1$, Stefano Laporta$^{4,5}$, Pierpaolo Mastrolia$^{4,5}$}
  \affiliation{
$^1$Istituto per le Applicazioni del Calcolo ``M. Picone,'' CNR, I-00185 Rome, Italy\\
$^2$INFN, Sezione di Roma Tre, I-00146 Rome, Italy\\
$^3$Institut des Hautes Etudes Scientifiques, 91440 Bures-sur-Yvette, France\\
$^4$Dipartimento di Fisica ed Astronomia, Universit\`a di Padova, Via Marzolo 8, 35131 Padova, Italy\\
$^5$INFN, Sezione di Padova, Via Marzolo 8, 35131 Padova, Italy
}

\date{\today}

\begin{abstract}
A recently introduced approach to the classical gravitational dynamics of binary systems involves intricate integrals (linked
to a combination of nonlocal-in-time interactions with iterated $\frac1r$-potential scattering) 
which have so far resisted attempts at their analytical evaluation.
By using computing techniques developed for the evaluation of multi-loop Feynman integrals (notably Harmonic Polylogarithms
and Mellin transform) we show how to analytically compute all the integrals entering the nonlocal-in-time contribution to the classical
 scattering angle at the sixth post-Newtonian accuracy, and at the seventh order in Newton's constant, $G$ (corresponding
to six-loop graphs in the diagrammatic representation of the classical scattering angle).
\end{abstract}

\maketitle

 \section{Introduction}
 The detection of the gravitational wave  signals emitted by compact
binary systems~\cite{LIGOScientific:2018mvr} has opened a new
path for investigating the structure of the Universe, and offers a
novel tool for studying the gravitational interaction. 
The full exploitation of this new observational tool poses, however, the theoretical challenge 
to model with improved accuracy the gravitational wave signals
emitted during the last orbits of coalescing black-hole binaries.

The latter theoretical challenge has recently motivated the construction of a new approach \cite{Bini:2019nra}
to the analytical description of the classical conservative dynamics of binary systems. The latter approach 
is based on a novel way of combining results from several theoretical formalisms, developed for studying the
gravitational potential within classical General Relativity (GR): post-Newtonian (PN) expansion, post-Minkowskian (PM) expansion,
multipolar-post-Minkowskian expansion, effective-field-theory, gravitational self-force approach, and effective one-body method. 
Another feature of the approach of Ref. \cite{Bini:2019nra} is to combine knowledge from gauge-invariant observables of {\it bound motions}, 
and from gauge-invariant observables of {\it scattering motions}.
In view of its characteristic multi-pronged nature,
we henceforth refer to the method of Ref. \cite{Bini:2019nra} as the Tutti Frutti (TF) method. 
 
The TF method has succeeded in pushing the state of the art to the sixth post-Newtonian (6PN) accuracy in the
conservative dynamics of binary systems \cite{Bini:2020wpo,Bini:2020nsb,Bini:2020hmy}. More precisely, the TF method
has determined the full structure of  two gauge-invariant characterizations of the 6PN-accurate dynamics:
the scattering angle $\chi$, and the radial action $I_r$, both being considered as functions of the total 
center-of-mass (c.m.) energy, $E=\sqrt{s}$, and of the total c.m. angular momentum, $J$. Both quantities are given
as  double expansions in  powers of the gravitational constant $G$ (PM expansion), and of the inverse 
velocity of light $1/c$ (PN expansion), each term of these expansions being a polynomial in the
symmetric mass-ratio $\nu=m_1m_2/(m_1+m_2)^2$. Most of the  $O(200)$  coefficients
entering the latter gauge-invariant characteristics of the 6PN dynamics have been analytically obtained
within the TF method except for six  coefficients entering  the local-in-time Hamiltonian.
In addition, the explicit implementation of the TF method requires the evaluation of a certain number
of  ``scattering integrals," $A_{m n k}$, arising in the computation of the nonlocal-in-time contribution to the scattering angle.
Previous work \cite{Bini:2020hmy} only succeeded in analytically computing a fraction of the latter 
scattering integrals: namely the $A_{m n k}$'s for $m=0,1$ and for $(mnk)=(200), (221)$.
 Some other scattering integrals (namely $A_{2 n k}$ for $(nk)=(20),(40),(41)$, $(42)$)
 were only numerically evaluated (with a modest, 8-digit accuracy).

Many computing techniques \cite{Caffo:1983tt,Laporta:1994yi,Vermaseren:1998uu,Remiddi:1999ew,Gehrmann:2001pz,Vollinga:2004sn,Maitre:2005uu,Huber:2007dx,Ablinger:2011te,Buehler:2011ev,Smirnov:2012gma,Panzer:2014caa,Duhr:2014woa,SL_2018,Blumlein:2018cms,Duhr:2019tlz,schnetz} have been developed for the evaluation of multi-loop Feynman integrals.  We show here how the use of some of these techniques,
notably involving the use of Mellin transforms \cite{Vermaseren:1998uu}, Harmonic Polylogarithms
(HPL) \cite{Remiddi:1999ew}, and expansion of hypergeometric functions about half-integer parameters \cite{Huber:2007dx},
allows one to derive the {\it analytical values} of all the scattering coefficients $A_{m n k}$'s 
entering the nonlocal-in-time contribution at the seventh order in $G$, and at the 6PN accuracy (the $G^7$ order
corresponds to the value $m=3$ of the first index $m$ of the scattering integrals $A_{m n k}$).
In particular, the present work will determine the exact, analytical values of the $O(G^6)$ scattering integrals $A_{2 n k}$
that were left undetermined in Ref. \cite{Bini:2020hmy}, and which enter the full determination of the 6PN {\it local-in-time}
dynamics, via the combination $D$, defined as 
(see  Eq. (6.29) of \cite{Bini:2020hmy})
\beq \label{DvsA2nk}
D = 
\frac1{\pi}\left(\frac52 A_{221}+\frac{15}{8}A_{200}+A_{242}\right)\,.
\eeq

The present work is an extension of Ref. \cite{Bini:2020uiq} which derived the analytical expressions
of the scattering coefficients $A_{2 n k}$ entering the nonlocal-in-time contribution at the sixth order in $G$.

\section{Setup on the GR side} \label{setup}

The TF method extracts information from various classical  GR observables. In particular, one of the
crucial gauge-invariant observables used in this approach is the {\it conservative}\footnote{See 
Refs. \cite{Bini:2012ji,Bini:2017wfr,DiVecchia:2020ymx,Damour:2020tta} for discussions including the radiation-reaction
contribution to the scattering angle.} classical scattering angle $\chi$ 
during a gravitational encounter,
considered as a function of the total c.m. energy, $E=\sqrt{s}$, the total c.m. angular momentum, $J$, and the symmetric mass
ratio $\nu$. We use the notation 
 \beq
 M \equiv m_1+m_2 \; ; \; \mu \equiv \frac{m_1 m_2}{m_1+m_2}\;;\; \nu \equiv \frac{\mu}{M} = \frac{m_1 m_2}{(m_1+m_2)^2}.
 \eeq
The TF approach  decomposes $\chi(E,J ;\nu)$ into three separate contributions: 
\beq \label{chitot0}
\chi(E,J,\nu)=\chi^{\rm loc, f}+ \chi^{\rm nonloc, h}+ \chi^{\rm f-h} \,,
\eeq
corresponding to an analogous decomposition of the total Hamiltonian:
$H(t)=H^{\rm loc,f}(t)+ H^{\rm nonloc,h}(t)+ \Delta^{\rm f-h}  H(t)$.
Here $\chi^{\rm loc, f}$  is the scattering angle that would be induced by the (f-route) local-in-time piece of the  Hamiltonian, 
$H^{\rm loc,f}(t)$. By contrast, $\chi^{\rm nonloc, h}$ is induced by the (h-route) nonlocal-in-time
piece of the Hamiltonian, $H^{\rm nonloc,h}(t)$,
while the last contribution, $\chi^{\rm f-h}$ is induced by the complementary (f-route) term $\Delta^{\rm f-h} H(t)$, which is
algorithmically derived \cite{Bini:2020hmy} from the $\nu$-structure of $\chi^{\rm nonloc, h}$. 
The present work will  focus on $\chi^{\rm nonloc, h}$, which is perturbatively determined as a double expansion
in  powers of the gravitational constant $G$ (PM expansion), and of the inverse velocity of light $1/c$ (PN expansion). 
It is convenient  to express the combined PM$+$PN expansion of  $\chi^{\rm nonloc, h}$
 in terms  of the dimensionless variables
\beq
\pinf \equiv \sqrt{\g^2-1} \ , \quad {\rm and \quad } j \equiv \frac{c J}{G m_1 m_2}\,,
\eeq
where the dimensionless energy parameter $\g $ is defined in terms of the total c.m. energy $E= \sqrt{s}$ by 
\beq
\g \equiv \frac{E^2-m_1^2c^4-m_2^2c^4}{2 m_1 m_2 c^4}\,.
\eeq
The variable $\g$ is equal both to the Lorentz factor between the two incoming worldlines, and  to the $\mu c^2$-rescaled  
 effective energy $\mathcal{E}_{\rm eff}$ entering the effective-one-body description \cite{Buonanno:1998gg} of the binary dynamics.
 
 As $j \propto \frac{c}{G}$, 
the PM expansion of  $\chi^{\rm nonloc, h}$ is equivalent to an expansion in inverse powers of $j$, and reads
\bea \label{chiexp}
&& \frac12 \chi^{\rm nonloc,h}(\g,j;\nu)=+ \nu \pinf^4\left(\frac{A_0^h(\pinf,\nu) }{ j^4}+ \frac{A_1^h(\pinf,\nu) }{ \pinf j^5}\right. \nonumber\\
&&\left.  \; \; \; \; +  \frac{A_2^h(\pinf,\nu) }{ \pinf^2 j^6}  +  \frac{A_3^h(\pinf,\nu) }{ \pinf^3 j^7} +O\left(\frac1{j^8} \right)\right).
\eea
The last-written contribution $\propto A_3^h(\pinf,\nu) /( \pinf^3 j^7)$ belongs to the 7PM approximation, $O(G^7)$.
The dimensionless coefficients $A_m^h(\pinf,\nu)$, $m=0,1,2,3,\cdots$, then admit a PN expansion, i.e., an expansion in powers of 
$\pinf =O\left(\frac1{c}\right)$, modulo logarithms of $\pinf$, say
\beq 
A_m^h(\pinf,\nu)= \sum_{n\geq0} \left[A_{m n}(\nu) + A^{\ln}_{m n}(\nu) \ln \left(\frac{p_{\infty}}{2} \right) \right] \pinf^n\,.
\eeq
The coefficient $A_{m n}(\nu)$ is a polynomial in $\nu$ of order $n$ and parametrizes a term of order 
$\frac{\pinf^{4+n-m}}{j^{4+m}} \sim\frac{ G^{4+m} }{c^{8+n}}$ (with $m \geq0$, $n \geq0$)
in the combined PM$+$PN expansion of the nonlocal scattering angle.
The leading-order contribution to the nonlocal dynamics is at the combined 4PM and 4PN level, i.e., $\propto G^4/c^8$ 
\cite{Blanchet:1987wq}. The corresponding nonlocal scattering coefficient, coming from $m=0$ and $n=0$, is  
$A_{0}^h(\pinf,\nu)= \pi\left[-\frac{37}{5}\ln \left(\frac{p_{\infty}}{2} \right)-\frac{63}{4}\right]+ O(\pinf^2)$ \cite{Bini:2017wfr}.
The higher-order logarithmic coefficients $A^{\ln}_{m n}(\nu)$ were analytically determined \cite{Bini:2020wpo,Bini:2020nsb,Bini:2020hmy}
so that we shall henceforth focus on the non-logarithmic coefficients $A_{m n}(\nu)$. 
Finally, the numerical scattering coefficient  $A_{m nk}$ is defined as the coefficient of the $k$th power of the
symmetric mass ratio $\nu$ in  $A_{m n}(\nu)$:
\beq
A_{m n}(\nu) \equiv \sum_{k=0}^{n} A_{m n k} \nu^k\,,
\eeq
with $k=0,1,2,\cdots$.

\section{Classical perturbative expansion of the nonlocal-in-time scattering angle}

Ref. \cite{Bini:2017wfr} has derived a general link (valid to first order in tail effects, i.e., up to 
$O\left[ (G^4/c^8)^2\right]= O\left[ G^8/c^{16}\right]$) between 
 the nonlocal-in-time contribution $\chi^{\rm nonloc, h}$ to the scattering angle and the integrated nonlocal action. Namely,
\beq \label{chinonloch}
\chi^{\rm nonloc,h}(E,J,\nu)= \frac{\partial W^{\rm nonloc,h}(E,J,\nu)}{\partial J}\,,
\eeq
where 
\beq \label{Wnonloc0}
W^{\rm nonloc,h}(E,J ; \nu) \equiv \int_{- \infty}^{+ \infty}dt\, H^{\rm nonloc,h}(t)\,,
\eeq
is the integrated (h-route) nonlocal action. The TF method expresses the latter quantity by the following explicit (regularized) two-fold integral
[to be evaluated along an hyperbolic-motion solution of the local-in-time Hamiltonian $H^{\rm loc,f}(t)$],
\beq \label{Wnonloch}
\!W^{\rm nonloc,h}=\!
 \alpha {\rm Pf}_{\Delta t^h}
\Int_{- \infty}^{+ \infty}  
\Int_{- \infty}^{+ \infty}
\frac{dt dt'}{|t-t'|} {\cal F}_{\rm GW}^{\rm split}(t,t') + \!\ O(\alpha^2)\,.
\eeq
Here: $\alpha \equiv G E/c^5=G \sqrt{s}/c^5$; ${\rm Pf}_h$
denotes the partie-finie regularization of the logarithmically divergent $t'$ integration at $t'=t$
(using the harmonic-coordinate-based time scale $\Delta t^h= 2 r_{12}^h(t)/c$);
 and ${\cal F}_{\rm GW}^{\rm split}(t,t')$ is the
time-split version (defined below) of the  gravitational-wave energy flux (absorbed and then) emitted  by the system\footnote{We consider
the conservative dynamics of a binary system interacting in a
time-symmetric way.}. The nonlocal expansion \eq{Wnonloch} is keyed by successive powers of $\alpha$.  
 The  $O(\alpha)$ term is called first-order tail; the $O(\alpha^2)$ is the second-order tail contribution, etc.
The effects linked to the second-order tail contribution have been analytically derived in \cite{Bini:2020hmy},
 at the combined 6PM and 5.5PN accuracy.
 [The next term in the PN expansion of the second-order tail contribution is at the 6.5PN level, 
which is beyond the accuracy sought for in the present work.]
 
We shall deal first with terms belonging to the $O(\alpha)$, first-order
 tail contribution explicated above. 
 The time-split version  of the  gravitational-wave energy flux is given, at the needed accuracy, by
 \begin{eqnarray}
\label{flux2PNdef}
&&{\cal F}_{ \rm 2PN}^{\rm split}(t,t')=\frac{G}{c^5} \left[ \frac15 I_{ab}^{\rm (3)}(t) I_{ab}^{\rm (3)}(t') \right. \nonumber\\
&& +\eta^2\left( \frac1{189 } I_{abc}^{\rm (4)}(t) I_{abc}^{\rm (4)}(t') +\frac{16}{45 } J_{ab}^{\rm (3)}(t) J_{ab}^{\rm (3)}(t')\right) \nonumber\\
&&+\eta^4 \left(\frac{1}{9072}I_{abcd}^{\rm (5)}(t) I_{abcd}^{\rm (5)}(t')
\left. +\frac{1}{84}J_{abc}^{\rm (4)}(t) J_{abc}^{\rm (4)}(t')\right)\right]\,.\nonumber\\
\end{eqnarray}
Here $\eta\equiv1/c$ and the superscript in parenthesis  indicates  repeated time-derivatives. The multipole moments $I_L$, $J_L$ denote
the values of the canonical moments $M_L$, $S_L$  entering the PN-matched~\cite{Blanchet:1987wq,Blanchet:1989ki,Damour:1990ji,Blanchet:1998in,Poujade:2001ie} multipolar-post-Minkowskian (MPM) formalism~\cite{Blanchet:1985sp}, when they are
reexpressed as explicit functionals of the instantaneous state of the binary system. These multipole moments parametrize
 (in a minimal, gauge-fixed way) the exterior gravitational field
(and therefore the relevant coupling between the system and a long-wavelength external radiation field). 

The subscript 2PN on ${\cal F}_{ \rm 2PN}^{\rm split}(t,t')$ indicates that
the multipole moments must be individually evaluated with the PN accuracy needed for knowing ${\cal F}_{ \rm 2PN}^{\rm split}(t,t')$,
and the corresponding ordinary (non time-split) gravitational wave flux,
\beq
{\cal F}_{ \rm 2PN}(t)={\cal F}_{ \rm 2PN}^{\rm split}(t,t)\,,
\eeq
with a {\it fractional} 2PN accuracy. More explicitly, this means that we need the 2PN-accurate value of the quadrupole moment expressed in 
terms of the material source~\cite{Blanchet:1995fr,Blanchet:1995fg}. The other moments (the electric octupole moment $I_{ijk}$, 
the electric hexadecapole moment, $I_{ijkl}$, the magnetic quadrupole moment, $J_{ij}$, and the magnetic octupole moment, $J_{ijk}$) 
need only to be known at the 1PN fractional accuracy \cite{Blanchet:1989ki,Damour:1990ji,Damour:1994pk}. Their explicit expressions 
(in the center-of-mass  harmonic coordinate frame) have been recalled in Eq. (3.3) and in Table I of Ref.  \cite{Bini:2020hmy}.

 Introducing the shorthand notation
\beq
\langle F \rangle_\infty  \equiv \int_{- \infty}^{+ \infty} dt F(t) \,,
\eeq
and expressing the partie-finie operation ${\rm Pf}_{\Delta t^h}$ entering Eq. \eq{Wnonloch} in terms of a partie-finie operation
 ${\rm Pf}_{2s/c}$ involving an intermediate length scale $s$, we decompose the nonlocal integrated action $W^{\rm nonloc,h}$
into two contributions
 \beq \label{W1W2}
W^{\rm nonloc,h}(E,j)= W_1^{\rm tail,h}(E,j) +W_2^{\rm tail,h}(E,j) + O(\alpha^2)\,,
\eeq
where
\bea \label{W1}
W_1^{\rm tail,h}(E,j)
&\equiv & - \alpha \left\langle {\rm Pf}_{2s/c}
\int_{-\infty}^\infty
 \frac{dt'}{|t-t'|}{\mathcal F}^{\rm split}_{\rm 2PN}(t,t') \right\rangle_\infty ,\nonumber\\
\eea
and
\bea 
\label{W2}
W_2^{\rm tail,h}(E,j) &\equiv&
2 \alpha  \left\langle {\mathcal F}_{\rm 2PN}(t)  \ln \left( \frac{r_{12}^h(t)}{s}\right) \right\rangle_\infty
.
\eea
 The integrated nonlocal action $W^{\rm nonloc,h}(E,j)$, and therefore each partial contribution, Eqs. \eq{W1}, \eq{W2}, has to be evaluated 
 along a 2PN-accurate hyperbolic motion.

\section{Quasi-Keplerian parametrization of the hyperbolic  motion, and its large-eccentricity expansion}

In view of Eq. \eq{chinonloch},  the PM expansion \eq{chiexp} of  $\chi^{\rm nonloc, h}$ is equivalent to the
following expansion of the integrated nonlocal action $W^{\rm nonloc,h}(E,j)$ in inverse powers of $j$, 
\bea \label{Wexp}
&&\frac{c \,W^{\rm nonloc,h}(\g,j;\nu)}{2\, G \, m_1 m_2}=- \nu \pinf^4\left(\frac{A_0^h(\pinf,\nu) }{3 j^3}+ \frac{A_1^h(\pinf,\nu) }{4 \pinf j^4}\right. \nonumber\\
&&\left. +  \frac{A_2^h(\pinf,\nu) }{5 \pinf^2 j^5} +  \frac{A_3^h(\pinf,\nu) }{6 \pinf^3 j^6} + O\left(\frac1{j^7} \right) \right)\,.
\eea
Remembering the proportionality between $j = cJ/(G m_1 m_2)$ and the impact parameter $b$ (via $J =b P_{\rm c.m.}$,
 where $P_{\rm c.m.}$ is the c.m. linear momentum of each body), we see that the computation of the scattering coefficients 
 $A_m^h(\pinf,\nu)$ amounts to expanding the integrated nonlocal action in inverse powers of $b$.
An explicit way to compute the large-impact-parameter expansion of  $W^{\rm nonloc,h}$ is to use the
 quasi-Keplerian parametrization \cite{DD85} of the 2PN-accurate hyperbolic-motion solution \cite{Cho:2018upo} 
of the 2PN dynamics of a binary system in harmonic coordinates \cite{DD1981a,D1982}.

The hyperbolic quasi-Keplerian parametrization involves a semi-major-axis-like quantity $a_r$, together with
several eccentricity-like quantities $e_t,e_r,e_\phi$. The variable parametrizing the time development
is an eccentric-anomaly-like (hyperbolic) angle $v$ varying from $-\infty$ to $+\infty$:
\begin{eqnarray} \label{hypQK2PN}
r&=& \bar a_r (e_r \cosh v-1)\,,\nonumber\\
\ell &=& \bar n (t-t_P)=e_t \sinh v-v + f_t V(v)+g_t \sin V(v)\,,\nonumber\\
\bar \phi &=&\frac{\phi-\phi_P}{K}=V(v)+f_\phi \sin 2V(v)+g_\phi \sin 3V(v)\,.\nonumber\\
\end{eqnarray}
Here, we use adimensionalized variables (and $c=1$), notably $r=r^{\rm phys}/(GM)$, $t= t^{\rm phys}/(GM)$,
while $V(v)$ is given by
\beq
\label{Vdef}
V(v)=2\, {\rm arctan}\left[ \Omega_{e_\phi}\tanh \frac{v}{2}  \right]\,,
\eeq
where
\beq
\Omega_{e_\phi} \equiv \sqrt{\frac{e_\phi+1}{e_\phi-1}}\,.
\eeq
The expressions (as functions of the specific binding energy $\bar E \equiv (E_{\rm tot}-Mc^2)/(\mu c^2)$ and of the dimensionless  
angular momentum $j=c J/(GM\mu)$) of the orbital parameters  $\bar n$ (hyperbolic mean motion) and $K$ (hyperbolic periastron precession), 
as well as $\bar a_r$, $e_t,e_r,e_\phi$, $f_t,g_t,f_\phi, g_\phi$, can be found in Appendix A of Ref. \cite{Bini:2020hmy}.
Let us only recall here the expressions of $\bar a_r$, and $e_r$ in terms of $\bar E$ and $j$:
\bea \label{arer}
\bar a_r &=& \frac{1}{2\bar E}\left[1-\frac{1}{2}\bar E \eta^2 (-7+\nu)\right. \nonumber\\
&&\left.+\frac{1}{4}\bar E^2\eta^4 \left(1+\nu^2-8 \frac{(-4+7\nu)}{\bar E j^2}\right)\right] 
\,,\nonumber\\
e_r^2 &=& 1+2\bar E j^2+\bar E [5\bar Ej^2(\nu-3)+2\nu-12]\eta^2\nonumber\\
&&
+\frac{\bar E}{j^2}[(4\nu^2+80-45\nu)\bar E^2j^4\nonumber\\
&&+(\nu^2+74\nu+30)\bar Ej^2+56\nu-32]\eta^4
\,.
\eea
When using this quasi-Keplerian parametrization, the combined PM$+$PN expansion of $W^{\rm nonloc,h}(\g,j;\nu)$
can be constructed from the combined large-$e_r$$+$large-$a_r$ expansion of the function $W^{\rm nonloc,h}(e_r,a_r)$.
On the one hand, as the tail action starts at the 4PN level, we need to work to the next-to-next-to-leading-order (NNLO) in 
$\frac1{a_r} \sim \frac{\pinf^2}{c^2}$ in order to reach the 6PN accuracy. On the other hand, as the tail action starts at the 4PM level 
($O(G^4)$),  we need to work to the next-to-next-to-next-to-leading-order (N$^3$LO) in 
$\frac1{e_r} $ in order to reach the 7PM, $O(G^7)$, accuracy (seventh order in $\frac1{b}$).

Without presenting too many technical details, let us illustrate the origin of some of the structures entering the scattering integrals $A_{mnk}$
by explaining how one can compute the large-eccentricity expansion  of the crucial nonlocal integral
\beq
\int \int dt dt' \frac{dt dt'}{|t-t'|} {\cal F}_{\rm GW}^{\rm split}(t,t')
\eeq
entering $W^{\rm nonloc,h}$. The first step is to introduce the auxiliary time variable $T \in [-1,1]$:
\beq
T \equiv\tanh \frac{v}{2}\,.
\eeq
In terms of this variable,  the 2PN-accurate functional relation between the original (rescaled) time variable
$t \equiv \frac{t^{\rm phys}}{G M}$ and the hyperbolic eccentric anomaly $v$  reads
\bea \label{tvsT}
 t&=& \frac{2}{\bar n}\left[ e_t \frac{T}{(1-T^2)}- {\rm arctanh}(T)\right.\nonumber\\
&+& f_t {\rm arctan}\left(\Omega_{e_\phi} T\right) 
\left.  
+g_t \frac{\Omega_{e_\phi}T}{1+\Omega^2_{e_\phi}T^2}
\right]\,,
\eea
with a corresponding expression for $t'$ vs $ T' \equiv \tanh \frac{v'}{2}$.
One then forms  $|t-t'|$, whose 2PN-accurate large-eccentricity expansion reads
\bea
|t-t'|&=&|T-T'|\frac{1+TT'}{(1-T^2)(1-T'^2)}\bar a_r^{3/2}e_r\nonumber\\
&&\times 
\left[2-(1+2\nu)\frac{\eta^2}{\bar a_r}+\frac14(8\nu^2-8\nu-1)\frac{\eta^4}{\bar a_r^2}\right]\nonumber\\
&&\times
\left[1+\frac{1}{e_r}{\mathcal P}_1+\frac{1}{e_r^2}{\mathcal P}_2  +\frac{1}{e_r^3}{\mathcal P}_3  +O\left(\frac1{e_r^4}\right)\right]\,,\nonumber\\
\eea
with coefficients ${\mathcal P}_1$, ${\mathcal P}_2$ and ${\mathcal P}_3$ of the form
\bea
{\mathcal P}_1&=&{\mathcal P}_{10}(T,T')+{\mathcal P}_{12}(T,T')\frac{\eta^2}{\bar a_r}+{\mathcal P}_{14}(T,T')\frac{\eta^4}{\bar a_r^2}\,, \nonumber\\
{\mathcal P}_2&=&{\mathcal P}_{24}(T,T')\frac{\eta^4}{\bar a_r^2}\,, \nonumber\\
{\mathcal P}_3&=&{\mathcal P}_{34}(T,T')\frac{\eta^4}{\bar a_r^2}\,.
\eea
Let us illustrate the structure of the coefficients ${\mathcal P}_{nm}(T,T')$ entering the ${\mathcal P}_n$'s by citing the expressions
of the first few of them. Introducing the shorthand notation
\bea \label{defkkappa}
At(T,T')&\equiv&{\rm arctan}(T)-{\rm arctan}(T')\,,\nonumber\\
Ath(T,T')&\equiv&{\rm arctanh}(T)-{\rm arctanh}(T')\,,
\eea
we have
\begin{widetext}
\begin{eqnarray}
{\mathcal P}_{10}(T,T')&=&-\frac{(1-T'^2)(1-T^2)}{(T T'+1)(T-T')} Ath(T,T')
\,,\nonumber\\
{\mathcal P}_{12}(T,T')&=&
\frac12(-8+3\nu){\mathcal P}_{10}(T,T')
\,,\nonumber\\
{\mathcal P}_{14}(T,T')&=& 
\frac18\nu(-29+3\nu){\mathcal P}_{10}(T,T')
+\frac{1}{8}\nu(-15+\nu)\frac{(T T'-1)(1-T'^2)(1-T^2)}{(1+T'^2)(1+T^2)(T T'+1)} 
\,,\nonumber\\
{\mathcal P}_{24}(T,T')&=& -\frac{3}{2}(-5+2\nu)\frac{(1-T'^2)(1-T^2)}{ (T T'+1) (T-T')} At(T,T')
+\frac12\nu(\nu-15)\frac{TT'(1-T^2)(1-T'^2)}{(1+T^2)^2(1+T'^2)^2}\nonumber\\
&&
-\frac18(16-43\nu+\nu^2)\frac{(1-T^2)^2(1-T'^2)^2}{(1+T^2)^2(1+T'^2)^2}
+2(-4+7\nu)\frac{(T^2T'^2+1)(T^2+T'^2)}{(1+T^2)^2(1+T'^2)^2}
\,,\nonumber\\
{\mathcal P}_{34}(T,T')&=&\frac32(-4+7\nu){\mathcal P}_{10}(T,T')
-\frac12\nu(\nu-15)\frac{TT'(TT'-1)(1-T'^2)(1-T^2)}{(TT'+1)(1+T^2)^2(1+T'^2)^2}\nonumber\\
&&
+\frac18(\nu^2+9\nu-60)\frac{(TT'-1)(1-T'^2)(1-T^2)[(TT'+1)^2-(T-T')^2][(TT'-1)^2-(T+T')^2]}{(TT'+1)(1+T^2)^3(1+T'^2)^3}\nonumber\\
&&
+6(-5+2\nu)\frac{(TT'-1)(1-T'^2)(1-T^2)[T^2(1+T'^2)^2+T'^2(1+T^2)^2]}{(TT'+1)(1+T^2)^3(1+T'^2)^3}
\,.
\end{eqnarray}
\end{widetext}
Using the above relations one can compute the large-eccentricity expansion of the measure
\beq
\frac{dt dt'}{|t-t'|}=\frac{1}{|t(T)-t'(T')|}\frac{dt}{dT}\frac{dt'}{dT'}dTdT' \equiv  d{\mathcal M}_{(T,T')}\,.
\eeq
Its schematic 2PN-accurate structure reads
\begin{eqnarray}
d{\mathcal M}_{(T,T')}&=& 2e_r\bar a_r^{3/2}\left[1-\frac{1+2\nu}{2\bar a_r}\eta^2-\frac{1+8\nu-8\nu^2}{8\bar a_r^2}\eta^4\right]\nonumber\\
&\times&
\frac{(1+T'^2) (1+T^2)dT dT'}{(1-T'^2)(1-T^2)(1+TT')|T-T'|} \nonumber\\
&\times& 
\left(1+\frac{{\mathcal M}_1}{e_r}+\frac{{\mathcal M}_2}{e_r^2}  +\frac{{\mathcal M}_3}{e_r^3}  +O\left(\frac1{e_r^4}\right) \right),\nonumber\\
\end{eqnarray}
where we have explicitly shown only the LO contribution in the large-eccentricity expansion. The  NLO, NNLO and N$^3$LO contributions
(respectively described by the coefficients ${\mathcal M}_1(T,T';\nu,\eta)$, ${\mathcal M}_2(T,T';\nu,\eta)$ and 
${\mathcal M}_3(T,T';\nu,\eta)$) have long expressions that we do
not explicitly display here. Let us simply note that  (recalling the definitions Eq. \eq{defkkappa})
${\mathcal M}_1(T,T';\nu,\eta)$ involves the function $Ath(T,T')$ linearly, 
${\mathcal M}_2(T,T';\nu,\eta)$ involves $Ath(T,T')$,  $Ath^2(T,T')$ and $At(T,T')$, while  
${\mathcal M}_3(T,T';\nu,\eta)$ involves $Ath(T,T')$,  $Ath^2(T,T')$,   $Ath^3(T,T')$ as well as  $At(T,T')$ and $Ath(T,T') At(T,T')$.

As illustrated here, apart from rational functions of $T$ and $T'$, the large-eccentricity expansion has a polynomial dependence
on the transcendental functions ${\rm arctan}(T)$,  ${\rm arctan}(T')$, ${\rm arctanh}(T)$ and ${\rm arctanh}(T')$.
Using these expansions (as well as corresponding expansions of the various multipole moments), one finally gets explicit
integral expressions for the scattering coefficients  $A_{m n k}$ 
of the form
\bea \label{Amnkint}
A_{m n k}  
&=&  \int_{- 1}^{+ 1}   \int_{- 1}^{+ 1}  \frac{dT dT'}{|T-T'|}a_{mnk}(T,T')\,,
\eea 
with integrands $a_{mnk}(T,T')$ of the form
\bea 
\label{amnk}
 a_{mnk}(T,T')=\sum_{p,q\geq 0}  
R_{pq}^{m n k}(T,T') Ath(T,T')^p At(T,T')^q\,,\nonumber\\
\eea
where $ R_{pq}^{m n k}(T,T')$ are  rational functions of $T$ and $T'$, and where we used the shorthands \eq{defkkappa}. 
The highest power of $Ath(T,T') \equiv{\rm arctanh}(T)-{\rm arctanh}(T')$ in this expression
is directly equal to the order of expansion in $\frac1{e_r}$ (and therefore in $G$, recalling 
the leading-order expression $e_r \approx \sqrt{1 + 2 {\bar E} j^2}$)
of the relativistic hyperbolic motion.

Ref. \cite{Bini:2020hmy}  succeeded in analytically computing  (up to the 6PN accuracy)
the numerical coefficients $A_{m n k}$ when $m=0$  ($G^4$ level) and  $m=1$ ($G^5$ level). 
By contrast, the integrands of Eq. \eq{Amnkint} become so involved when $m=2$ and $m=3$ 
 ($G^6$ and $G^7$ levels) that most of them resisted analytical integration by standard integration
methods. 

\section{Multiple polylogarithms and harmonic polylogarithms}
\label{sec:hpl}
To determine the analytic expressions of the scattering integrals $A_{2nk}$
we follow one of the strategies used in  the realm of multi-loop Feynman
calculus, namely the reduction to {\it iterated integrals} \cite{Caffo:1983tt,Laporta:1994yi,Remiddi:1999ew,Gehrmann:2001pz,Vollinga:2004sn,Maitre:2005uu,Ablinger:2011te,Buehler:2011ev,Smirnov:2012gma,Panzer:2014caa,Duhr:2014woa,SL_2018,Blumlein:2018cms,Duhr:2019tlz,schnetz}. 
Given a sequence of univariate functions 
$g_{a_1}(x), g_{a_2}(x), \cdots,g_{a_n}(x)$, assumed (say) to be regular at $x=0$,
iterated integrals are recursively defined by 
$  G(a_1,a_2,\cdots,a_n;x)= \int_0^x dt_1 g_{a_1}(t_1) G(a_2,\cdots,a_n;t_1)$, with the starting value $G(\emptyset;x)=1$.  
The simplest class of iterated integrals are the {\it multiple polylogarithms}
defined by considering a sequence of inverse-linear functions: $g_{a_i}(x)= (x-a_i)^{-1}$. These were introduced by Poincar\'e \cite{Poincare1884}, and have been the topic of many mathematical studies, e.g., \cite{Goncharov:1998kja,Borwein:1999js,Brown:2009qja,Deligne:2010,Schnetz:2017bko}. They also came up as important tools for expressing certain multi-loop Feynman integrals
\cite{Broadhurst:1996kc,Panzer:2014caa}. 
On the other hand, from the practical
point of view, a {\it subclass} of the multiple logarithms, the harmonic polylogarithms (HPL) \cite{Remiddi:1999ew}, has turned out to be 
sufficient, and very useful, to express many Feynman integrals. They are defined by restricting the singular points $a_i$ entering 
$  G(a_1,a_2,\cdots,a_n;x)$ to taking one of the three values $+1, -1$ or $0$, and by normalizing the inverse-linear factors in
a slightly different way. Specifically, the HPLs are defined as the recursive integrals,
\beq
H_{i_1, i_2, {\ldots} i_n}(x) = 
\int_0^x     {dt_1}\; 
f_{i_1}(t_1)
H_{i_2 {\ldots} i_n}(t_1)  
\ ,
\eeq
with $f_{\pm 1}(x)=(1\mp x)^{-1}$, $f_0(x)=1/x$, and a regularization at $x=0$ such that 
$H_{0, 0, {\ldots}, 0}(x) \equiv \ln^n (x)/n!$.

A crucial feature of the multiple polylogarithms, and therefore of the HPLs, is that they enjoy special algebraic properties,
going under the names of: shuffle algebra, stuffle algebra, scaling invariance, shuffle-antipode relations, H\"older convolution, 
integration-by-parts identities, etc.
In addition, all these special algebraic properties respect a filtration by the {\it weight}, i.e. by the number $n$ 
of singular values, $a_1,a_2,\cdots,a_n$, or the number $n$ of indices on $H_{i_1 i_2 {\ldots} i_n}(x)$. The weight
corresponds to the number of iterations appearing in the nested integral representation. For instance, at weight 1 a multiple
polylogarithm is a simple logarithm, while at weight 2, it is a linear combination of a dilogarithm and a squared logarithm.
The remarkable algebraic properties of  multiple polylogarithms (and HPLs) allow one to express them algebraically, at any given weight $n$,
in terms of a minimal subset of  them, having weights $n'\leq n$.  For instance, at weights $n =2, 3$, and $4$ the minimal subsets are
formed by 3, 8, and 18 elements, respectively.
In addition, their evaluation for special values of their 
arguments $a_1,a_2,\cdots,a_n;x$ can often be reduced to a relatively small number of transcendental constants. This is particularly
the case if, besides 0, the arguments $a_1,a_2,\cdots,a_n;x$ are roots of unity. For introductions to the vast literature on
the properties, and evaluation, of multiple polylogarithms and HPLs (including computer-program implementations) see, e.g.,
\cite{Goncharov:1998kja,Borwein:1999js,Gehrmann:2001pz, Vollinga:2004sn, Maitre:2005uu,Ablinger:2011te, Buehler:2011ev, Panzer:2014caa,Duhr:2014woa, Henn:2015sem,SL_2018,Duhr:2019tlz,Brown:2020rda,schnetz}.

\section{Analytic evaluation of the $O(G^6)$ scattering integrals via harmonic polylogarithms}
\label{sec:analyticeval}
Let us now sketch how we could analytically compute the $O(G^6)$ scattering integrals, i.e. Eq. \eq{Amnkint}, with $m=2$, 
by reducing these two-fold definite integrals to the evaluation of HPLs, of weight $\leq 4$, for the values $x=1,i$ of the HPL variable.

First, using symmetry properties of the integrands $a_{mnk}(T,T')$ entering Eq. \eq{amnk}, it is possible 
to reduce the double integration to the triangle $0<T<1, 0<T'<T$. Let us start by discussing
the integration over $T'$ on the interval $0<T'<T$. 
The crucial information needed for discussing this first integration concerns  
 the structure of the integrands $a_{mnk}(T,T')$, and particularly of the denominators entering the rational coefficients 
 $ R_{pq}^{m n k}(T,T')$ in Eq. \eq{amnk}, when $m=2$. 
 To be concrete, let us discuss the case $(mnk)=(242)$ and exhibit one representative part of the 
 integrand $a_{242}(T,T')$. It reads 
 \bea \label{Q42sample}
 && \frac{-16 (1-T^2)^3 (1-T'^2)^3 P_2(T,T') }{315 (1 + T^2)^8 (1 + T'^2)^8( 1+T T')^3 (T-T')^3} \nonumber\\
 && \times \left\{[{\rm arctanh}(T)-{\rm arctanh}(T')]^2 \right. \nonumber\\
&& \left.-\frac{(T-T')^2}{2(1-T^2)^2}  -\frac{(T-T')^2}{2 (1-T'^2)^2}  \right\},
 \eea
where $P_2(T,T')$ is a (symmetric) polynomial in $T$ and $T'$, of order 14 in both variables.
By partial fractioning \eq{Q42sample} with respect to $T'$ (keeping $T$ fixed) one is reduced to evaluating integrals of the type
\beq
\int dT' \frac{{\rm arctanh}^p(T')}{(T'-a)^q}\,,
\eeq
where $p=0,1,2$, $1\leq q \leq 8$ and $a=\pm i, -\frac1T, T$ or $\pm 1$. Integrating by parts (with respect to $T'$), one can reduce
the power $q$ down to $q=1$. At this stage, remembering that ${\rm arctanh}(T) = \frac12 {\rm ln}((1+T)/(1-T))$ (and  ${\rm arctan}(T)={\rm arctanh}(i T)/i$ for other denominators) are (as explained above) of weight 1, we see that the highest-weight term in the numerator,
$\propto \ln^2((1+T')/(1-T'))$, is of weight 2, so that its integration over $T'$ with the additional kernel $(T'-a)^{-1}$ will generate
terms of weight 3. The explicit computation of the needed integration over $T'\in [0,T]$, with the values of $a$ listed above,
 is found to involve at most the trilogarithm
${\mathrm{Li}}_3(z)$ at the rational arguments $z=- \frac{1+T}{1-T}$ or $z=- \left(\frac{1+T}{1-T}\right)^2$. 

Having so obtained an explicit weight-3 expression for the result of the integration over $T'$, we need to perform the 
final integration over $ T \in {[0,1]}$. This is done in three steps. 
The first step is the same that was used for the
$T'$ integration. There are now polynomial denominators  involving powers
of $T^2+1 $, powers of $T \pm 1$, and also powers of $T$. 
Partial-fractioning, and integrating by parts, one can reduce these powers to the first power. 
Second, we use the definition of HPLs to express the  integrals containing $T^{-1}$ and $(T \pm 1)^{-1}$ in terms of HPLs.
Third, we consider the integrals containing $(T \pm i)^{-1}$:
these cannot be directly cast in HPL format (which admits poles only at $T=0, \pm1$).
Therefore, we modify the integrands by the  insertion of a parameter $x$, to
be later replaced by a suitable value, so as to obtain the original integral back. 
Following a technique introduced many years ago to analytically evaluate multi-loop Feynman integrals \cite{Caffo:1983tt,Laporta:1994yi},
the integral, now function of $x$, is reduced to iterated integrals of the type $\int_0^x dx_1 (x_1-a_1)^{-1} \int_0^{x_1} dx_2 (x_2-a_2)^{-1}\cdots$, 
by combining repeated differentiations with respect to $x$ with partial-fractioning, and integrations by parts,
followed by quadratures to get back the original integral.

Let us show an example of this technique: all the $A_{2nk}$ integrals contain, after the $T'$ integration, the same 
combination of integrals of weight $w=4$,
\beq\label{jex1}
\hat J= \int_0^1 dT \frac{-2 \, \ln^3\left(\frac{1-T}{1+T}\right) -3 \,{\mathrm{Li}}_3\left[
-\left(\frac{1-T}{1+T}\right)^2\right]}{1 + T^2} \ .
\eeq
We modify the integral \eq{jex1}, to let it acquire a dependence on
the new parameter $x$, {i.e.} $\hat J \to J(x)$, in the following way:
\bea\label{jex2}
&& J(x)\equiv
i \int_0^1 dT \, (1-x^2) \times \nonumber \\
&& 
\times  \frac{-2 \, \ln^3\left(\frac{1-T}{1+T}\right)  -3 \,{\mathrm{Li}}_3\left[
\left(\frac{(1-T)(1-x)}{(1+T)(1+x)}\right)^2\right]}{2 x (T+x)(T+1/x)}.
\eea
It is easily seen that the original integral is recovered at the value $x=i$, that is $\hat J=J(i)$,
and that $J(1)=0$.
By differentiating and reintegrating three times over $x$,
on the model of
  $J(x)=\int_1^x dx (dJ(x)/dx)$,
  $J(x)$ can be expressed in terms of HPLs at weight $w\leq 4$, namely: 
\bea\label{jexhpl2}
&& i \, J(x)
= \frac{23}{240}\pi^4 -  21 \ln (2) \,\zeta(3) + \pi^2\ln^2 (2) - \ln^4 (2)
- 24 a_4
\nonumber\\
&&
+ \frac{21}{2} H_{-1}(x) \zeta(3)
- \frac{3}{2} H_{0}(x) \zeta(3)
+ \frac{21}{2} H_{1}(x) \zeta(3)
\nonumber\\
&&
+\frac{1}{2} \pi^2 H_{0,-1}(x)
+\frac{1}{2} \pi^2 H_{0,1}(x)
-\frac{3}{2} \pi^2 H_{-1, -1}(x)
\nonumber\\
&&
-\frac{3}{2} \pi^2 H_{-1, 1}(x)
-\frac{3}{2} \pi^2 H_{1, -1}(x)
-\frac{3}{2} \pi^2 H_{1, 1}(x) 
\nonumber\\
&&
+ 12 H_{0,1, -1}(x) \ln (2)
+ 12 H_{0,1, 1}(x) \ln (2)
\nonumber\\
&&
- 12 H_{0,-1, -1, -1}(x)
+  6 H_{0,-1, -1, 0}(x)
- 12 H_{0,-1, 1, -1}(x)
\nonumber\\
&&
+  6 H_{0,-1, 1, 0}(x)
- 12 H_{0,1,-1, -1}(x)
+  6 H_{0,1, -1, 0}(x)
\nonumber\\
&&
- 12 H_{0,1, 1, -1}(x)
+  6 H_{0,1, 1, 0}(x)
-  6 H_{-1, -1, -1, 0}(x)
\nonumber\\
&&
-  6 H_{-1, -1, 1, 0}(x)
-  6 H_{-1, 1, -1, 0}(x)
-  6 H_{-1, 1, 1, 0}(x) 
\nonumber\\
&&
-  6 H_{1, -1, -1, 0}(x)
-  6 H_{1, -1, 1, 0}(x)
-  6 H_{1, 1, -1, 0}(x)
\nonumber\\
&&
-  6 H_{1, 1, 1, 0}(x)
+ 12 H_{0,-1, -1}(x) \ln (2)
+ 12 H_{0,-1, 1}(x) \ln (2) \ . \nonumber \\
\eea
This result  expresses $\hat J=J(i) $ in terms of the values at the fourth root of unity, $i$,  of HPLs of weight $w\leq 4$ (together with 
$a_4 \equiv {\rm Li}_4(1/2)$, and lower-weight quantities such as $\pi^2$ and $\zeta(3)$).
Using \cite{Maitre:2005uu}, we expressed the needed values of the HPLs at $x=i$
in terms of a small subset of irreducible constants of weight $w\leq4$, namely: 
$ \catd ={\rm Im Li}_2 (i) =\sum_{n=0}^\infty (-1)^n/(2n+1)^2$ (Catalan's constant), 
$ \Q3={\rm Im}H_{0,1,1}(i) $, $ \Q4={\rm Im}H_{0,1,1,1}(i) $, $ a_4={\rm Li}_4(1/2)$ and $ \cat4={\rm Im Li}_4 (i) $.
The irreducible weight-4 constants are found to cancel when evaluating
$\hat J=J(i) $ by means of Eq. \eq{jexhpl2} to yield
\beq
\hat J=J(i) = -\frac{1}{2}\pi^2\; {\rm{K}} + \frac{9}{2} \pi
\zeta(3)\,.
\eeq

Applying our technique to all the scattering integrals $A_{2nk}$, we found that they could all be expressed in terms of
the values of HPLs of weight $w\leq4$ at the arguments $x=1$ or $x=i$. Similarly to what happens for $\hat J=J(i) $,
the irreducible weight-4 constants are found to cancel in the evaluation of all the scattering integrals $A_{2nk}$.  
Actually, the final results for the $A_{2nk}$'s
are found to factorize as the product of $\pi$ with constants of weight $\leq 3$. For instance, we found
\beq
 A_{242}  = - \pi \left(\frac{583751}{864}+ \frac{100935}{64}\, \zeta(3) \right)\,.
 \eeq
Our complete analytical results for the $A_{2nk}$'s are listed in Table \ref{A3nk_res}.
We give below the relations between such coefficients and those used in Ref. \cite{Bini:2020hmy} to parametrize the 
 (non-logarithmic) part of the scattering angle (see Eq. (4.15) there)
\bea \label{Ankvsdnk}
\pi^{-1} A_{200} &=& d_{00}, \nonumber\\
\pi^{-1} A_{220} &=& d_{20}+3 d_{00}, \nonumber\\
\pi^{-1} A_{221} &=& d_{21}-2 d_{00}, \nonumber\\
\pi^{-1} A_{240}  &=& d_{20}+d_{40}+\frac32 d_{00}, \nonumber\\
\pi^{-1} A_{241}  &=& d_{21}-\frac{11}{2}d_{00}+d_{41}-2d_{20}, \nonumber\\
\pi^{-1} A_{242}  &=& d_{42}-2d_{21}+3d_{00}\,.
\eea
Further details about our integration procedures, and our intermediate results, are provided in the Supplemental Material \cite{SupMat}.


\begin{table}
\caption{\label{A2nk_res} Analytical results for the $O(G^6)$ scattering coefficients $A_{2nk}$.}
\begin{ruledtabular}
\begin{tabular}{ll}
coefficient& value\\
\hline
$\pi^{-1} A_{200}$&$ -\frac{99}{4}-\frac{2079}{8}\, \zeta(3)$ \\
$\pi^{-1} A_{220}$&$ -\frac{41297}{112}-\frac{9216}{7}\ln(2)+\frac{49941}{64}\, \zeta(3)$ \\
$\pi^{-1} A_{221}$&$ \frac{1937}{8}+\frac{3303}{4}\, \zeta(3)$ \\
$\pi^{-1} A_{240}$&$ \frac{1033549}{4536}+\frac{10704}{7}\ln(2)-\frac{40711}{128}\, \zeta(3)$ \\
$\pi^{-1} A_{241}$&$ \frac{8008171}{8064}+\frac{75520}{21}\ln(2)-\frac{660675}{256}\, \zeta(3)$ \\
$\pi^{-1} A_{242}$&$ -\frac{583751}{864}-\frac{100935}{64}\, \zeta(3)$\\
\end{tabular}
\end{ruledtabular}
\end{table}

\section{Evaluation of the $O(G^7)$ scattering integrals}
\label{sec:evalG7}

At the $O(G^7)$ level, i.e., for the integrals $A_{mnk}$ with index $m=3$, the structure of the integrands $a_{mnk}(T,T')$
becomes more complex. The rational functions $R_{pq}^{m n k}(T,T')$ entering as coefficients in Eq. \eq{amnk} involve
higher-order polynomials in their numerators, but their most important feature, namely the location of the poles in the denominators, 
stays the same as at the $O(G^6)$ level. Again the poles are located at $T = T'$, $T= -1/T'$, $T = \pm 1$, $T = \pm i$, 
$T' = \pm 1$ and $T'=\pm i$.
However, an important change concerns the powers $p$ and $q$ with which the functions 
$At(T,T')\equiv{\rm arctan}(T)-{\rm arctan}(T')$ and $Ath(T,T')\equiv{\rm arctanh}(T)-{\rm arctanh}(T')$ enter the
numerator of $a_{mnk}(T,T')$. At the $O(G^7)$ level, we have the values $(p,q)= (0,0), (1,0), (2,0), (3,0), (0,1), (1,1)$.
In particular, the highest value of $p+q$ is 3, and is reached via the presence of a term proportional to $Ath^3(T,T')$.
We already noticed that both $Ath(T,T')$ and $ At(T,T')$ are of weight 1. The integrand $R_{pq}^{3 n k}(T,T') Ath^3(T,T')$
is therefore of weight 3. Its double integral over $T$ and $T'$ can therefore be {\it a priori} expected to be of weight 5.

We succeeded in finding the analytic expressions of the $O(G^7)$ integrals entering the 6PN nonlocal scattering
angle by using several methods. As a preliminary method, we combined very-high-precision (200 digits)
numerical computation of the integrals (using a double-exponential change of variables \cite{DoubleExponential})
 with the PSLQ algorithm \cite{PSLQ} and a basis of  transcendental constants indicated by the structure of the integrands.
 Let us recall that such an {\it experimental mathematics} strategy is often used in the realm of multi-loop Feynman
calculus, when a direct analytic integration seems prohibitive, see
{\it e.g.}, Refs. \cite{Broadhurst:1996kc,Laporta:2017okg,Laporta:2019fmy}.
Previous uses of experimental mathematics and high-precision arithmetics within studies of binary systems
include Refs. \cite{Shah:2013uya,Johnson-McDaniel:2015vva,Foffa:2016rgu}. We note in passing that 
one of  the integrals (in momentum space) contributing to the 4PN-static term of the two-body potential, used in
\cite{Foffa:2016rgu} and originally obtained by analytic recognition \cite{Lee:2015eva},
was later analytically confirmed by direct integration (in position space) \cite{Damour:2017ced}.

The application of experimental mathematics to the  $A_{3nk}$ integrals has shown that, similarly to what happened
for the $A_{2nk}$ integrals, the final results were simpler than what was {\it a priori} expected. In particular, 
we found that the final results do not go beyond weight 4, and that the only weight-4 quantity entering (some
of) the results is simply $\zeta(4) \propto \pi^4$. 

Having obtained such simple semi-analytic expressions for the $A_{3nk}$ integrals,
we embarked on confirming them by means of a purely analytical derivation. 
We found that an efficient method for doing so was to reformulate the time-domain
integral defining the integrated action Eq. \eq{Wnonloch} in frequency space. 
Actually, when decomposing $W^{\rm nonloc,h}(E,j)$ in the two contributions
entering Eq. \eq{W1W2}, the most difficult one to evaluate is
\bea \label{W1bis}
W_1^{\rm tail,h}(E,j)
&\equiv & - \alpha \left\langle {\rm Pf}_{2s/c}
\int_{-\infty}^\infty
 \frac{dt'}{|t-t'|}{\mathcal F}^{\rm split}_{\rm 2PN}(t,t') \right\rangle_\infty\,.\nonumber\\
\eea
In order to express $W_1^{\rm tail,h}(E,j)$ in the frequency domain, the first step is 
to Fourier transform\footnote{In the following, we use $GM=1$, i.e., we work with $GM$-rescaled
time and frequency variables.} the multipole moments. For example,
\beq
\label{I_ab_fourier}
I_{ab}(t)=\int_{-\infty}^{+\infty} \frac{d\omega}{2\pi}e^{-i \omega t}\hat I_{ab}(\omega)\,,
\eeq
where
\beq
\label{I_ab_omega}
\hat I_{ab}(\omega)=\int_{-\infty}^{+\infty} dt e^{i\omega t}I_{ab}(t)\,,
\eeq
with the associated PN expansion 
\bea
\label{PN_Iab}
\hat I_{ab}(\omega)&=&\hat I_{ab}^{\rm N}(\omega)+\eta^2\hat I_{ab}^{\rm 1PN}(\omega)\nonumber\\
&+&\eta^4\hat I_{ab}^{\rm 2PN}(\omega)
+O(\eta^6)\,.
\eea
Inserting these Fourier representations into Eq. \eqref{W1bis} then yields (see Section V of Ref. \cite{Bini:2017wfr} for details)
\begin{eqnarray}
\label{W1_final_exp}
W_1^{\rm tail,h}(E,j)&=& 2\frac{G^2H_{\rm tot}}{\pi c^5} \int_0^\infty d\omega {\mathcal K}(\omega) \ln\left(\omega\frac{2s}{c}e^\gamma\right)
\,,\nonumber\\
\end{eqnarray}
where 
\begin{eqnarray}
{\mathcal K}(\omega)&=&\frac15 \omega^6 |\hat I_{ab}(\omega)|^2\nonumber\\
&+&\eta^2 \left[\frac{\omega^8}{189}|\hat I_{abc}(\omega)|^2+\frac{16}{45}\omega^6 |\hat J_{ab}(\omega)|^2  \right]\nonumber\\
&+&\eta^4 \left[\frac{\omega^{10}}{9072}|\hat I_{abcd}(\omega)|^2+\frac{\omega^8}{84}|\hat J_{abc}(\omega)|^2  \right]\,,\nonumber\\
\end{eqnarray}
and we have used the result 
\beq
{\rm Pf}_T \int_0^\infty d\tau \frac{\cos \omega \tau}{\tau}=-\ln (|\omega| T e^\gamma)\,,
\eeq
with $\gamma=0.577215\ldots$. 
Note the close link between the expression  \eqref{W1_final_exp} for $W_1^{\rm tail,h}(E,j)$ and
the frequency-domain expression of  the total energy flux emitted during the scattering process, namely
\beq \label{EGWomega}
\Delta E_{\rm GW}=\frac{G}{\pi c^5} \int_{0}^\infty  d\omega  {\mathcal K}(\omega)\,.
\eeq
The difference between the two expressions is embodied in the logarithmic factor $\ln\left(\omega\frac{2s}{c}e^\gamma\right)$, 
which is characteristic of the tail in the frequency domain \cite{Blanchet:1993ec}.

The relation between $\Delta E_{\rm GW}$ and $W_1^{\rm tail,h}(E,j)$ is clarified by stating them in the
framework of the Mellin transform. Let us first note that it is convenient to replace the frequency $\omega$
by the variable $u$, using
\beq \label{defu}
\omega \equiv \frac{u}{e_r \bar a_r^{3/2}} \,,
\eeq
so that Eq. \eqref{W1_final_exp} becomes\footnote{We take here $e_r$ and $\bar a_r$, as fundamental variables. At any stage of the calculation
these can be re-expressed, via Eq. \eq{arer}, as functions of $E$ and $j$.}
\begin{eqnarray}
\label{W1_fourier}
W_1^{\rm tail,h}(E,j)&=& 2\ln \alpha_s G H_{\rm tot}\Delta E_{\rm GW}\nonumber\\
&+&
2\frac{G^2H_{\rm tot}}{\pi c^5}\frac{1}{e_r \bar a_r^{3/2}} \int_0^\infty du \mathcal K(u) \ln u \,,\nonumber\\
\end{eqnarray}
where
\beq \label{EGWu}
\Delta E_{\rm GW}=\frac{G}{\pi c^5}\frac{1}{e_r \bar a_r^{3/2}}  \int_{0}^\infty  du  {\mathcal K}(u)\,,
\eeq
with
\beq
{\mathcal K}(u)={\mathcal K}(\omega)\big|_{\omega= u/(e_r \bar a_r^{3/2})}\,,
\eeq
and
\beq
\alpha_s =  \frac{2s}{ce_r \bar a_r^{3/2}}e^\gamma \,.
\eeq
We recall that the Mellin transform of a function $f(u)$ (with $u \in [0,+\infty]$) is defined as 
\beq
\label{mellin}
g(s)\equiv \mathfrak M\{f(u);s\}=\int_0^\infty u^{s-1} f(u) du\,.
\eeq
It is then easily seen  that the first two terms of the Taylor expansion of $g(s)$ around $s=1$
are respectively given by
\beq
g(1)= \int_0^\infty f(u) du\,,
\eeq
and
 \beq
\frac{dg(s)}{ds}\bigg|_{s=1}=\int_0^\infty  f(u)\ln u du\,.
\eeq
This shows the possible usefulness of the Mellin transform in connecting $W_1^{\rm tail,h}$ to $\Delta E_{\rm GW}$.

We have indeed been able to use the Mellin transform to analytically compute all the scattering integrals at $O(G^7)$,
i.e. the values of the integrals appearing in the double PN+PM (or $\eta^2$-$e_r^{-1}$) expansion of
\beq
 \int_0^\infty du \ln u \left[\mathcal K(u)\right]^{\rm PN+PM} \,,
\eeq
where
\bea
\label{eq_718}
&&\left[\mathcal K(u)\right]^{\rm PN+PM}= {\mathcal K}_{\rm N}^{\rm LO}(u) +  \eta^2{\mathcal K}_{\rm 1PN}^{\rm LO}(u)  +  \eta^4{\mathcal K}_{\rm 2PN}^{\rm LO}(u) \nonumber\\
&&+ \frac{1}{e_r} {\mathcal K}_{\rm N}^{\rm NLO}(u) + \frac{\eta^2}{e_r} {\mathcal K}_{\rm 1PN}^{\rm NLO}(u)+   \frac{\eta^4}{e_r} {\mathcal K}_{\rm 2PN}^{\rm NLO}(u)\nonumber\\
&&+ \frac{1}{e_r^2} {\mathcal K}_{\rm N}^{\rm NNLO}(u) + \frac{\eta^2}{e_r^2} {\mathcal K}_{\rm 1PN}^{\rm NNLO}(u)+   \frac{\eta^4}{e_r^2} {\mathcal K}_{\rm 2PN}^{\rm NNLO}(u)\nonumber\\
&&+ \frac{1}{e_r^3} {\mathcal K}_{\rm N}^{\rm N^3LO}(u) + \frac{\eta^2}{e_r^3} {\mathcal K}_{\rm 1PN}^{\rm N^3LO}(u)+   \frac{\eta^4}{e_r^3} {\mathcal K}_{\rm 2PN}^{\rm N^3LO}(u),
\eea
as well as their simpler analogs appearing in the double $\eta^2$-$e_r^{-1}$ expansion of
\beq
 \int_0^\infty du  \left[\mathcal K(u)\right]^{\rm PN+PM} \,.
\eeq
The starting point of this approach rests on the simple value of the Fourier transform of the multipole moments
at the lowest PN order, i.e. at the Newtonian order ($O(\eta^0)$), but at all orders in $\frac1{e_r}$:
\beq
\left[\mathcal K(u)\right]_{\rm N}={\mathcal K}_{\rm N}^{\rm LO}(u) + \frac{1}{e_r} {\mathcal K}_{\rm N}^{\rm NLO}(u) + \frac{1}{e_r^2} {\mathcal K}_{\rm N}^{\rm NNLO}(u) + \cdots
\eeq 
In the elliptic-motion case, it is well-known that the (discrete) Fourier expansion of the Newtonian multipole moments 
involve ordinary Bessel functions,
namely $J_{p+k}( p e_r)$, where $p$ and $k$ are integers. In the  hyperbolic-motion case the (continuous) Fourier transform
of the Newtonian-level multipole moments involve integrals of the form
\beq
\int_{-\infty}^\infty e^{q\sinh v -(p+k) v}dv= 2e^{-i \frac{\pi}{2}(p+k)}K_{p+k}(u)\,,
\eeq
involving the modified Bessel function $K_{p+k}(u)$ of real argument $u$, Eq. \eq{defu}, but of order $p+k$, where $k=0,\pm1, \cdots$ is
an integer, while $p$ defined as
\beq
p \equiv \frac{q}{e_r }\,, \qquad 
q \equiv i \, u\,, 
\eeq
is purely imaginary, and $u$-dependent.
The Newtonian-level energy integrand $\left[\mathcal K(u)\right]_{\rm N}$ is quadratic in time-derivatives of the
Newtonian multipole moments. Remembering the fact that the variable $u$ is proportional to the frequency,  
$\left[\mathcal K(u)\right]_{\rm N}$ therefore involves functions of the type
\beq \label{k1k2k3}
u^{k_1}  K_{p+k_2}(u)  K_{p+k_3}(u)\,,
\eeq
with some integers $k_1, k_2, k_3$.

There are several technical features which allow one to compute integrals involving bilinear quantities in 
Bessel $K$ functions of the type \eq{k1k2k3}. First, the Mellin transform $g_{\rm KK}(s; \mu, \nu)$ 
of the function $ f_{\rm KK}(u;\mu,\nu)\equiv K_\mu(u)K_\nu(u)$
has a simple explicit expression, namely
\bea  \label{gkk}
g_{\rm KK}(s; \mu, \nu) &=& \frac{ 2^{s-3}}{\Gamma(s)}\nonumber\\
&\times& \Gamma\left(\frac{s+\mu+\nu}{2} \right)\Gamma\left(\frac{s-\mu+\nu}{2} \right)\nonumber\\
&\times &
\Gamma\left(\frac{s+\mu-\nu}{2} \right)\Gamma\left(\frac{s-\mu-\nu}{2} \right).
\eea
Differentiating the result \eq{gkk} with respect to the Mellin parameter $s$ then allows one to compute the
$\ln u$-weighted integral of  integrands of the form \eq{k1k2k3}.

The situation becomes more involved when going beyond the Newtonian level. Indeed,  
the post-Newtonian-level Fourier-domain {\it integrands}  
${\mathcal K}_{\rm 1PN}^{\rm LO}(u) $, ${\mathcal K}_{\rm 1PN}^{\rm NLO}(u) $, etc
can no longer be explicitly computed. For instance, the 1PN-level, $\frac1{e_r}$-NLO term $\mathcal K_{\rm 1PN}^{\rm NLO}(u)$
reads
\begin{widetext}
\bea \label{K1pnnlo}
\mathcal K_{\rm 1PN}^{\rm NLO}(u)&=&\frac{16}{21}u^3\left[
\left(u^4-46 u^2-\frac{141}{5}\right)K_0(u)^2+\frac{122}{5}u\left(u^2-\frac{653}{122}\right)K_0(u)K_1(u)+\left(u^4-\frac{333 u^2}{10}-\frac{39}{5}\right)K_1(u)^2
\right]\nonumber\\
&&
-\frac{48}{5\pi}u^4\int_{-\infty}^{\infty} dv\, {\rm arctan}\left(\tanh\frac{v}{2}\right)\left[
\sinh 2v(K_0(u)+2uK_1(u))\cos(u\sinh v)\right.\nonumber\\
&&\left.
+\frac12(\cosh 3v-5\cosh v)(uK_0(u)+K_1(u))\sin(u\sinh v)
\right]\nonumber\\
&&
-\frac{64}{21}u^3\left[
\left(u^4-\frac{21 u^2}{20}-\frac{3}{4}\right)K_0(u)^2-\frac{6}{5}u\left(u^2+\frac{95}{24}\right)K_0(u)K_1(u)+\left(u^4-\frac{23 u^2}{20}-\frac{21}{20}\right)K_1(u)^2
\right]\nu
\,.\nonumber\\
\eea
\end{widetext}
Here, the terms on the second and third lines involve an integral over $v$ which cannot be explicitly evaluated.
[This $v$-integral comes from the original integral
$\int dt e^{ i \omega t} I_{ab\cdots}(t)= \int dv \frac{dt(v)}{dv}  I_{ab\cdots}(t(v))$ defining the Fourier-transformed 
multipole moments, when inserting for the function $t(v)$ the PN+eccentricity expansion of the relativistic
Kepler equation, see Eq. \eq{hypQK2PN}, which notably involves the (large-eccentricity-expanded) function $V(v)$, Eq. \eq{Vdef}.]
However, it is still possible to analytically evaluate the resulting double integral $\int du \ln u \int dv [\cdots] $ 
by integrating first over $u$ (using Mellin-transform properties to replace the $\ln u$ factor by an $s$-derivative),
and then integrating over $v$. These computations could be done because we could 
 obtain explicit expressions for the Mellin transforms $g_{\rm Kcos}(s;v,\nu)$ and $g_{\rm Ksin}(s;v,\nu)$ of the functions 
\beq
f_{\rm Kcos}(u;v,\nu) \equiv  K_\nu(u)\cos(u\sinh v)\,,
\eeq
 and 
 \beq
 f_{\rm Ksin}(u;v,\nu) \equiv K_\nu(u)\sin(u\sinh v)\,,
 \eeq
which appear in Eq. \eq{K1pnnlo}, and also at higher PN orders. See Eqs. \eq{gdefs}.

Last, but not least, we are interested in expanding the integrals in the large eccentricity limit, $e_r \to \infty$.
In this limit, the $u$-dependent order $p = \frac{i u}{e_r }$ tends to zero, so that the $\frac1{e_r}$ expansion 
is equivalent to evaluating derivatives with respect to the {\it order}, $\nu$, of Bessel $K_{\nu}$ functions.

Using all those technical features of the frequency-domain integrals
(as well as the programme \texttt{HypExp2} \cite{Huber:2007dx}, which allows one to
evaluate the Taylor-expansion of hypergeometric functions around half-integer values of their parameters,
see Eq. \eq{dhypergeom}), 
we were able to derive  analytic expressions 
for all the scattering coefficients $A_{3nk}$ (which confirmed the results previously obtained by experimental
mathematics techniques). More technical details on our analytical derivations are given in Appendix \ref{app_fourier}.
The final results for the N$^3$LO scattering coefficients $A_{3nk}$ appearing at the 6PN level are listed
in Table \ref{A3nk_res}.


\begin{table}
\caption{\label{A3nk_res} Analytical results for the $O(G^7)$ scattering coefficients $A_{3nk}$.} 
\begin{ruledtabular}
\begin{tabular}{ll}
coefficient& value\\
\hline
$A_{300}$&$ \frac{79936}{225}-\frac{18688}{15}\ln(2)-\frac{88576}{75}\zeta(3)$ \\
$A_{320}$&$ \frac{2239456}{1575}-\frac{568448}{105}\ln(2)-\frac{64256}{525}\zeta(3)-\frac{621}{20}\pi^4$ \\
$A_{321}$&$ -\frac{384}{175}+\frac{96512}{15}\ln(2)+\frac{901632}{175}\zeta(3)$ \\
$A_{340}$&$ \frac{57597448}{51975}+\frac{1175968}{567}\ln(2)+\frac{135861232}{17325}\zeta(3)$\\
&$-\frac{16848}{25}\pi^2+\frac{31779}{448}\pi^4$ \\
$A_{341}$&$ -\frac{1677767408}{259875}+\frac{22912832}{1575}\ln(2)-\frac{12013696}{3465}\zeta(3)$\\
&$+\frac{14067}{112}\pi^4$ \\
$A_{342}$&$ -\frac{5455648}{2205}-\frac{237824}{15}\ln(2)-\frac{132771328}{11025}\zeta(3)$ \\
\end{tabular}
\end{ruledtabular}
\end{table}

\section{Final results for the nonlocal contributions to the scattering angle at $O(G^7)$}

As briefly recalled in Sec. \ref{setup}, there are three types of contributions to the scattering angle,
as displayed in Eq. \eq{chitot0}: the f-route local contribution  $\chi_n^{\rm loc, f}$, the h-route nonlocal contribution $\chi^{\rm nonloc, h}$,
and the additional contribution $\chi^{\rm f-h}$. The f-route local contribution, $\chi_n^{\rm loc, f}$ was computed  (at the 6PN accuracy)  
up to $G^7$ included in  Ref. \cite{Bini:2020nsb} (see Eq. (8.2) there). The two remaining contributions are related to nonlocal effects. 
Previous results \cite{Bini:2019nra,Bini:2020wpo,Bini:2020nsb,Bini:2020hmy} 
on $\chi_6^{\rm nonloc, h}$ and $\chi_6^{\rm f-h}$ were complete only up to order $G^5$. Here, we shall give complete results 
up to order $G^7$, within the 6PN accuracy.

The h-route nonlocal contribution $\chi^{\rm nonloc, h}$, in Eq. \eq{chitot0}, is directly linked (via Eq. \eq{chinonloch})
to the integrated nonlocal action $W^{\rm nonloc,h}(E,J ; \nu)$, Eqs. \eq{Wnonloc0}, \eq{Wnonloch}. The work done in the sections
above has allowed us to derive the analytical expressions of the expansion coefficients $A_{mnk}$ parametrizing $W^{\rm nonloc,h}(E,J ; \nu)$,
and therefore $\chi^{\rm nonloc, h}$ (as per Eq. \eq{chiexp}). We gather the final results for the function $\chi^{\rm nonloc, h}(\g,j,\nu)$
in the following subsection. 

The last contribution, $\chi^{\rm f-h}$,  in Eq. \eq{chitot0} to the scattering angle is indirectly related to nonlocal effects.
As explained in Refs. \cite{Bini:2020wpo,Bini:2020nsb,Bini:2020hmy}, this additional contribution is defined as
\beq
\frac12 \chi^{\rm f-h}(\g,j;\nu) = \frac1{2 M^2 \nu} \frac{\partial W^{\rm f-h}(\g,j;\nu)}{\partial j}\,,
\eeq
where $W^{\rm f-h}(\g,j;\nu)$ is the  additional contribution to the integrated action
related to the use of a suitably flexed partie-finie scale $ f(t) \Delta t^h= 2 f(t) r_{12}^h(t)/c$
in the definition of the nonlocal Hamiltonian. This generates the following result for $W^{\rm f-h}$:
\begin{eqnarray}
\label{Wf1}
W^{\rm f-h}&=& + 2\frac{G H_{\rm tot}}{c^{5}}\int dt {\mathcal F}^{\rm split}_{\rm 2PN}(t,t)  \ln(f(t))\,.
\end{eqnarray}
As discussed in Refs. \cite{Bini:2020wpo,Bini:2020nsb,Bini:2020hmy}, the flexibility factor is determined, modulo
some gauge freedom, by the few contributions to the function $\chi^{\rm nonloc, h}(\g,j,\nu)$ that violate the
simple $\nu$-dependence rules \cite{Damour:2019lcq} satisfied by the total scattering angle $\chi^{\rm tot}$.
The resulting value of $\chi^{\rm f-h}(\g,j;\nu)$ will be discussed in the second subsection below.

Before listing our results for the various contributions to the scattering angle, let us recall our conventional
definition of the expansion coefficients in the large-$j$ limit (which include a factor $\frac12$):
\beq
\frac12 \chi^{\rm tot}(\g,j;\nu)=\sum_{n\geq1}\frac{\chi_n(\g,\nu)}{j^n}\,,
\eeq
with
\beq
\chi_n(\g,\nu)= \chi_n^{\rm loc, f}(\g;\nu)+ \chi_n^{\rm nonloc, f}(\g;\nu)\,.
\eeq
The various pieces of the nonlocal part  
\beq
\chi_n^{\rm nonloc, f}(\g;\nu)=\chi_n^{\rm nonloc, h}(\g;\nu)+ \chi_n^{\rm f-h}(\g;\nu)\,,
\eeq
with
\beq
\chi_n^{\rm nonloc, h}= \chi_n^{\rm h, \alpha}+ \chi_n^{\rm h, \alpha^2} + O(\alpha^3)\,,
\eeq
will be shown as a 6PN-accurate expansion (keyed by the powers of $\pinf \equiv \sqrt{\g^2-1}$) of the type
\beq
\chi_n^{\rm h, \alpha}= \chi_n^{\rm h, \alpha, 4PN} +  \chi_n^{\rm h, \alpha, 5PN}+  \chi_n^{\rm h, \alpha, 6PN}\,,
\eeq
and
\beq
\chi_n^{\rm h, \alpha^2}=\chi_n^{\rm h, \alpha^2, 5.5PN}\,.
\eeq
Note that the third-order-tail contribution starts at the 7PN level, which is beyond the PN accuracy sought for in the present work.

\subsection{The h-route first-order-tail contribution to the scattering angle}

The $\frac1j$-expansion coefficients of the $4+5+6$PN contribution to the first-order-tail part of the scattering angle are given by
\begin{widetext}

\begin{eqnarray}
\pi^{-1}\chi_4^{\rm h, \alpha, 4PN}&=&
\left[-\frac{37}{5}\ln \left(\frac{p_{\infty}}{2} \right)-\frac{63}{4}\right]\nu p_{\infty}^4
\,,\nonumber\\
\pi^{-1}\chi_4^{\rm h, \alpha, 5PN}&=&
\left[\left(-\frac{1357}{280}+\frac{111}{10}\nu\right)\ln \left(\frac{p_{\infty}}{2} \right)-\frac{2753}{1120}+\frac{1071}{40}\nu\right]\nu p_{\infty}^6
\,,\nonumber\\
\pi^{-1}\chi_4^{\rm h, \alpha, 6PN}&=&
\left[\left(-\frac{27953}{3360}+\frac{2517}{560}\nu-\frac{111}{8}\nu^2\right
)\ln \left(\frac{p_{\infty}}{2} \right)-\frac{155473}{8960}+\frac{109559}{40320}\nu-\frac{186317}{5040}\nu^2\right]\nu p_{\infty}^8
\,,
\end{eqnarray}
\begin{eqnarray}
\chi_5^{\rm h, \alpha, 4PN}&=&
 \left[-\frac{6656}{45}-\frac{6272}{45}\ln \left(4\frac{p_{\infty}}{2} \right)\right]\nu p_{\infty}^3
\,,\nonumber\\
\chi_5^{\rm h, \alpha, 5PN}&=&
\left[\left(-\frac{74432}{525}+\frac{13952}{45}\nu\right)\ln \left(4\frac{p_{\infty}}{2} \right)+\frac{114368}{1125}+\frac{221504}{525}\nu\right]\nu p_{\infty}^5
\,,\nonumber\\
\chi_5^{\rm h, \alpha, 6PN}&=&
\left[\left(-\frac{881392}{11025}+\frac{288224}{1575}\nu-\frac{21632}{45}\nu^2\right)\ln \left(4\frac{p_{\infty}}{2} \right)
+\frac{48497312}{231525}-\frac{5134816}{23625}\nu-\frac{25465952}{33075}\nu^2\right]\nu p_{\infty}^7
\,,
\end{eqnarray}
\begin{eqnarray}
\pi^{-1}\chi_6^{\rm h, \alpha, 4PN}&=&
\left[ -122 \ln \left(\frac{p_{\infty}}{2} \right)-\frac{99}{4}-\frac{2079}{8}\zeta(3)
\right]\nu p_{\infty}^2\,,\nonumber\\
\pi^{-1}\chi_6^{\rm h, \alpha, 5PN}&=& 
\left[\left(\frac{811}{2}\nu-\frac{13831}{56}\right)\ln\left(\frac{p_{\infty}}{2} \right)
-\frac{41297}{112}-\frac{9216}{7}\ln(2)+\frac{49941}{64}\zeta(3)+\left(\frac{3303}{4}\zeta(3)+\frac{1937}{8}\right)\nu \right]\nu p_{\infty}^4
\,,\nonumber\\
\pi^{-1}\chi_6^{\rm h, \alpha, 6PN}&=& 
\left[\left(\frac{64579}{1008}-785\nu^2+\frac{75595}{168}\nu\right)\ln\left(\frac{p_{\infty}}{2} \right)
-\frac{40711}{128}\zeta(3)+\frac{1033549}{4536}+\frac{10704}{7}\ln(2)\right.\nonumber\\
&&\left.
+\left(\frac{75520}{21}\ln(2)+\frac{8008171}{8064}-\frac{660675}{256}\zeta(3)\right)\nu+\left(-\frac{100935}{64}\zeta(3)-\frac{583751}{864}\right)\nu^2
\right]\nu p_{\infty}^6
\,,
\end{eqnarray}
and 
\begin{eqnarray}
\chi_7^{\rm h, \alpha, 4PN}&=&
\left[-\frac{9344}{15}\ln \left(4\frac{p_{\infty}}{2} \right)+\frac{79936}{225}-\frac{88576}{75}\zeta(3)\right]\nu p_{\infty}
\,,\nonumber\\
\chi_7^{\rm h, \alpha, 5PN}&=&
\left[\left(-\frac{284224}{105}+\frac{48256}{15}\nu\right)\ln \left(4\frac{p_{\infty}}{2} \right)
-\frac{621}{20}\pi^4+\frac{2239456}{1575}-\frac{64256}{525}\zeta(3)+\left(\frac{901632}{175}\zeta(3)-\frac{384}{175}\right)\nu\right]\nu p_{\infty}^3
\,,\nonumber\\
\chi_7^{\rm h, \alpha, 6PN}&=&
\left[\left(-\frac{118912}{15}\nu^2+\frac{11456416}{1575}\nu+\frac{587984}{567}\right)\ln \left(4\frac{p_{\infty}}{2} \right)
+\frac{135861232}{17325}\zeta(3)+\frac{57597448}{51975}-\frac{16848}{25}\pi^2+\frac{31779}{448}\pi^4\right.\nonumber\\
&&\left.
+\left(-\frac{12013696}{3465}\zeta(3)+\frac{14067}{112}\pi^4-\frac{1677767408}{259875}\right)\nu
+\left(-\frac{5455648}{2205}-\frac{132771328}{11025}\zeta(3)\right)\nu^2\right]\nu p_{\infty}^5
\,.
\end{eqnarray}

\end{widetext}

\subsection{The h-route  second-order-tail contribution to the scattering angle}

The second-order-tail contribution $W_{5.5{\rm PN}}^{\rm nonloc}$ to the nonlocal integrated action is given by
\beq
\label{W5p5nldef}
W_{5.5{\rm PN}}^{\rm nonloc}=\alpha^2\left\langle \frac{B}{2}\int_{-\infty}^\infty
\frac{d\tau}{\tau}{\mathcal H}^{\rm split}(t,\tau) \right\rangle_{\infty}\,,
\eeq
where $B=-\frac{107}{105}$ and 
\beq
{\mathcal H}^{\rm split}(t,\tau)=\frac{G}{5c^5}[I^{(3)}_{ij}(t)I^{(4)}_{ij}(t+\tau)-I^{(3)}_{ij}(t)I^{(4)}_{ij}(t-\tau)]\,.
\eeq
Working in the Fourier domain we find
\begin{eqnarray}
W_{5.5{\rm PN}}^{\rm nonloc}&=&-\alpha^2 B \frac{G}{5c^5}\int_0^\infty
d\omega \omega^7 |\hat I_{ij}(\omega)|^2\,,\nonumber\\
\end{eqnarray}
where we have used the result 
\beq
\int_{-\infty}^\infty d\tau \frac{\sin\omega\tau}{\tau}=\pi\,.
\eeq
At our present level of accuracy, it is enough to use the Newtonian approximation to the Fourier transform $\hat I_{ij}(\omega)$ of the
quadrupole moment. Using the relations given in the previous section we have then
\beq \label{W55fourier}
W_{5.5{\rm PN}}^{\rm nonloc}=\alpha^2\frac{107}{105} \frac{G}{c^5} \frac{1}{e_r^2 \bar a_r^3}\int _0^\infty du\, u\,  {\mathcal K}_{\rm N}(u)\,.
\eeq

Using the results of Section 5 in \cite{Bini:2020hmy},  extending the large-eccentricity expansion to
the NNLO order and using the frequency-domain integrals presented in Appendix \ref{app_fourier}, one finds
\bea
\chi_5^{\rm h, \alpha^2, 5.5PN}&=& 
-\frac{47936}{675} \nu\pinf^6
\,,\nonumber\\
\pi^{-1}\chi_6^{\rm h, \alpha^2, 5.5PN}&=&
-\frac{10593}{560}\pi^2 \nu\pinf^5
\,,\nonumber\\
\chi_7^{\rm h, \alpha^2, 5.5PN}&=&
\left(\frac{499904}{1575}+\frac{4738816}{23625}\pi^2\right) \nu\pinf^4
\,.\nonumber\\
\eea

\subsection{The f-h additional contribution to the scattering angle}

The flexibility factor $f(t)$ has been determined in terms of the 6PN-accurate, $O(G^6)$ h-route nonlocal
scattering angle in Section VII of \cite{Bini:2020hmy}. 
[Our new results at the $O(G^7)$ level do not change the determination of the flexibility factor.] 
The corresponding additional contribution
\beq \label{Hfmh0}
\Delta^{\rm f-h}  H= 2 \frac{G H_{\rm tot}}{c^5} {\mathcal F}^{\rm split}_{\rm 2PN}(t,t)  \ln(f(t))\,,
\eeq 
to the f-route nonlocal Hamiltonian has been determined in \cite{Bini:2020hmy} (Eq. (7.29) there)
to be equal, modulo an irrelevant canonical transformation, to
\beq \label{decompHf-h}
\Delta^{\rm f-h} H'_{\rm 5+6PN}=\Delta^{\rm f-h}{ H'}^{\rm min}_{\rm 5+6PN}+ \Delta^{\rm f-h} {H'}^{C D}_{\rm 5+6PN}\,.
\eeq
Here,  $\Delta^{\rm f-h}{ H'}^{\rm min}_{\rm 5+6PN}$ denotes the {\it minimal} part of  the canonically-transformed $\Delta^{\rm f-h}  H$
(built with the minimal solution, Eq. (7.28) there), while
$\Delta^{\rm f-h} {H'}^{C D}_{\rm 5+6PN}$ denotes the part that involves six arbitrary flexibility parameters, namely:  
 $C_2$, $C_3$, $D_2^0$, $D_3^0$, and $D_4= D_4^0+ \nu D_4^1$. Explicitly, the latter contribution reads
\bea \label{Hfmenh6PNunconstr}
&&\frac{\Delta^{\rm f-h} {H'}^{C D}_{\rm 5+6PN}}{M}= C_2 \frac{\nu^3 p_r^2}{r^5} +  C_3\frac{\nu^3}{r^6}\nonumber\\
&&
+\left(D_2^0+ \frac{14}{3} \nu C_2\right)\frac{\nu^3 p_r^4}{r^5}
\nonumber\\
&&
+\left[ D_3^0+ \nu\left(-\frac{3}{2} C_2+6C_3\right)\right]\frac{\nu^3 p_r^2}{r^6}+(D_4^0+\nu D_4^1)\frac{\nu^3}{r^7}
\,.\nonumber\\
\eea 
On the other hand, the fully determined minimal Hamiltonian  $\Delta^{\rm f-h}{ H'}^{\rm min}_{\rm 5+6PN}$
given in Eq. (7.30) of \cite{Bini:2020hmy} involves the coefficient
\beq \label{DvsA2nkbis}
D = \frac1{\pi}\left(\frac52 A_{221}+\frac{15}{8}A_{200}+A_{242}\right)\,,
\eeq
which could not be analytically determined in \cite{Bini:2020hmy}. Our new results, presented above, allow
one to determine the exact analytical expression of the coefficient $D$. Though the individual scattering coefficients $A_{2nk}$
entering $D$ involve $\zeta(3)$, it is remarkably found that $D$
turns out to be equal to the rational number
\beq
D= -\frac{12607}{108}\,,
\eeq
which is compatible with the previous numerical estimate of \cite{Bini:2020hmy}, namely 
$D^{\rm num}= -116.73148147(1)$. The value of $D$ then determines the minimal value of the flexibility coefficient
$D_3^{\rm min}$ (see Eq. (7.28) in \cite{Bini:2020hmy}), namely 
\bea \label{D3min}
D_3^{\rm min}  = -\frac{68108}{945}\nu \,,
\eea
as well as the $f$-related, 6PN-level contribution to the periastron precession (see Eq. (8.30) in \cite{Bini:2020hmy}):
\begin{eqnarray}
K^{\rm f-h,  circ, min}(j)&=& + \frac{68108}{945}\frac{\nu^3}{ j^{12}} \,.
\end{eqnarray}
Inserting the analytical value of $D$ in Eq. (7.30) of \cite{Bini:2020hmy} also determines the analytical value of 
$\Delta^{\rm f-h}{ H'}^{\rm min}_{\rm 5+6PN}$, namely  
\bea \label{Hfmenh6PNnewmin}
\frac{\Delta^{\rm f-h}{ H'}^{\rm min}_{\rm 5+6PN}}{M} &=&\nu^3\frac{168}{5} \frac{p_r^4}{r^4}
+\nu^3\left(\frac{271066}{4725} + \frac{21736}{189}\nu \right) \frac{p_r^6}{r^4} \nonumber\\
 &-& \nu^4 \frac{39712}{189} \frac{p_r^4}{r^5}-
 \nu^4 \frac{68108}{945}\frac{p_r^2}{r^6}\,.\nonumber\\
\eea

Using the (canonically transformed) additional Hamiltonian \eq{decompHf-h}, it is a straightforward matter to compute
the large-eccentricity expansion of the corresponding integrated action 
\begin{eqnarray}
\label{Wf1n}
W^{\rm f-h}&=& + 2\frac{G H_{\rm tot}}{c^{5}}\int dt {\mathcal F}^{\rm split}_{\rm 2PN}(t,t)  \ln(f(t)) \nonumber\\
&=& \int dt \Delta^{\rm f-h}{ H'}_{\rm 5+6PN}\,,
\end{eqnarray}
and the corresponding (halved) scattering angle contribution 
\bea
\frac12 \chi^{\rm f-h} = \frac1{2 M^2 \nu} \frac{\partial W^{\rm f-h}(\g,j;\nu)}{\partial j}\,.
\eea
We find
\begin{widetext}
\bea
\pi^{-1}\chi_4^{\rm f-h}&=&
-\frac{3}{32} C_1 \nu^2p_\infty^6
+\left(\frac{27}{64} C_1 \nu -\frac{3}{64} C_1-\frac{15}{256}D_1\right)\nu^2p_\infty^8
\,,\nonumber\\ 
\chi_5^{\rm f-h}&=&
\left(-\frac{8
   }{5}C_1-\frac{8 }{15}C_2\right)\nu^2p_\infty^5 
	+\left[\left(\frac{276}{35} C_1+\frac{32
   }{15}C_2\right)\nu  -\frac{172 }{35}C_1-\frac{4}{15} C_2-\frac{8
   }{7}D_1-\frac{8 }{35}D_2\right]\nu^2p_\infty^7 
\,,\nonumber\\ 
\pi^{-1}\chi_6^{\rm f-h}&=&
\left(-\frac{45
   }{32}C_1-\frac{15}{16} C_2-\frac{15}{16} C_3\right)\nu^2p_\infty^4\nonumber\\
	&+&
	\left[\left(\frac{495 }{64}C_1+\frac{275}{64} C_2+\frac{105
   }{32}C_3\right)\nu  -\frac{615 }{64}C_1-\frac{95
   }{32}C_2-\frac{15 }{32}C_3-\frac{75 }{64}D_1-\frac{15
   }{32}D_2-\frac{5}{32} D_3\right]\nu^2p_\infty^6 
\,,\nonumber\\ 
\chi_7^{\rm f-h}&=&
(-8 C_1-8 C_2-16
   C_3)\nu^2p_\infty^3\nonumber\\
	&+&  
	\left[\left(\frac{252 }{5}C_1+\frac{216 }{5}C_2+\frac{344
   }{5}C_3\right)\nu  -100 C_1-\frac{292}{5} C_2-\frac{264
   }{5}C_3-8 D_1-\frac{24 }{5}D_2-\frac{16
   }{5}D_3-\frac{16 }{5}D_4\right]\nu^2p_\infty^5 
\,,\nonumber\\
&&\\
\eea
\end{widetext}
with minimal values (for vanishing values of $C_2$, $C_3$, $D_2^0$, $D_3^0$, and $D_4= D_4^0+ \nu D_4^1$,
and the minimal values $C_1^{\rm min}, D_1^{\rm min}, D_2^{\rm min}, D_3^{\rm min}$ given in Eq. (7.28) 
in \cite{Bini:2020hmy}, with Eq. \eq{D3min} above)
\bea
\pi^{-1}\chi^{\rm f-h}_{4\,\rm min}&=&
-\frac{63}{20}\nu^2p_\infty^6
+\left(-\frac{199037}{40320}+\frac{27331}{10080}\nu\right)\nu^2p_\infty^8
\,,\nonumber\\ 
\chi^{\rm f-h}_{5\,\rm min}&=&
-\frac{1344}{25}\nu^2p_\infty^5\nonumber\\
&&+\left(-\frac{7629872}{33075}+\frac{2448608}{33075}\nu\right)\nu^2p_\infty^7 
\,,\nonumber\\ 
\pi^{-1}\chi^{\rm f-h}_{6\,\rm min}&=&
-\frac{189}{4}\nu^2p_\infty^4\nonumber\\
&&+\left(-\frac{786449}{2016}+\frac{12607}{108}\nu\right)\nu^2p_\infty^6 
\,,\nonumber\\ 
\chi^{\rm f-h}_{7\,\rm min}&=&
-\frac{1344}{5}\nu^2p_\infty^3\nonumber\\
&& +\left(-\frac{18044528}{4725}+\frac{90464}{75}\nu\right)\nu^2p_\infty^5 
\,.
\eea

\section{Conclusions}
 By using  computing techniques developed for the evaluation
 of  multi-loop Feynman integrals, we have advanced the analytical  knowledge of classical  gravitational scattering 
  at the seventh order in $G$, and at the sixth post-Newtonian  accuracy, by fully determining the {\it nonlocal-in-time}
  contribution to the scattering angle. The present work has given a new instance of a fruitful synergy between 
  classical GR and QFT techniques leading to an  improved  theoretical description
 of gravitationally interacting binary systems.

\section*{Acknowledgments}
DB and TD thank Massimo Bernaschi for collaboration at an early stage of this project, and for informative discussions on numerical integration. 
DB and PM acknowledge the hospitality, and the highly stimulating environment, of the Institut des Hautes Etudes Scientifiques. 
 SL thanks CloudVeneto for the use of computing and storage facilities. DB and AG thank MaplesoftTM for providing a complimentary license of Maple 2020.

\appendix

\section{Details on the frequency domain computation}
\label{app_fourier}

The first step is to Fourier transform\footnote{In the following, we use $GM=1$, i.e., we work with $GM$-rescaled
time and frequency variables.} the multipolar moments (see, e.g., Eq. \eqref{I_ab_omega}).
At the Newtonian level the computation is done by using the integral representation of the Hankel functions of the first kind of order $p$ and argument $q$ 
\beq
\label{Hankel_rep}
H_p^{(1)}(q)=\frac{1}{ i\pi }\int_{-\infty}^\infty e^{q\sinh v -p v}dv\,.
\eeq
As the argument $q=iu$  of the Hankel function is purely imaginary, the Hankel function becomes converted into a 
Bessel $K$ function, according to the  relation
\beq
H_p^{(1)}(iu)=\frac{2}{\pi}e^{-i \frac{\pi}{2}(p+1)}K_p(u)\,.
\eeq
Note that the order $p=i u/e_r$ of the Bessel functions is purely imaginary, and proportional to the (frequency-dependent)
argument $u = \omega e_r \bar a_r^{3/2}$. However, the order $p$ tends to zero when $e_r \to \infty$, which allows
most integrals to be explicitly computed when performing a large-eccentricity expansion.
A typical term at the Newtonian level ($O(\eta^0)$) is of the kind $e^{q\sinh v -(p+k) v}$, the Fourier transform of which is
\beq
e^{q\sinh v -(p+k) v}\to2e^{-i \frac{\pi}{2}(p+k)}K_{p+k}(u)\,,
\eeq
involving Bessel functions having the same argument $u$, but various orders differing by integers.
However,  standard identities valid for Bessel functions allow  one to reduce the orders $p+k$ to either $p$ or $p+1$. 
When taking the large-eccentricity expansion, one  expands with respect to the order of the Bessel functions.
This gives rise, at LO, to $K_0(u)$, and $K_1(u)$,  and at NLO, NNLO, N$^3$NLO, to derivatives of $K_0(u)$, and $K_1(u)$
with respect to their orders. 

Higher orders in the PN expansion ($O(\eta^2), O(\eta^4)$) imply for the integration in $v$ more complicated expressions like $v^ne^{q\sinh v -(p+k) v}$ and $e^{q\sinh v -(p+k) v}V(v)$. The Fourier transform of $v^ne^{q\sinh v -(p+k) v}$ leads to integrands involving
\beq
v^ne^{q\sinh v -(p+k) v}\to2(-1)^n\frac{\partial^n}{\partial p^n}\left[e^{-i \frac{\pi}{2}(p+k)}K_{p+k}(u)\right]\,,
\eeq
while the Fourier transform of the terms $e^{q\sinh v -(p+k) v}V(v)$ requires to work with
 the large-$e_r$ expansion of the $V$-term (see Eq. \eqref{Vdef}), i.e.,
\begin{eqnarray}
\label{arctan_exp}
V(v)&=&2\,{\rm arctan}\left(\tanh \frac{v}{2}\right)
\nonumber\\
&+&
\frac1{e_r}\tanh v + \frac{\sinh v}{e_r^2 \cosh^2 v}
+O(e_r^{-3})\,.
\end{eqnarray}
One then generally has terms of the form $e^{q\sinh v -(p+k) v}f_j(v)$, involving non-trivial functions $f_j(v)$,
which cannot be integrated analytically.
However, in most cases one can overcome this difficulty by integrating over $u$, before integrating over $v$.

\subsection{Integrating over the frequency spectrum and Mellin transform}
 
The integrated nonlocal action $W_1^{\rm tail,h}$ (Eq. \eqref{W1_fourier})  and the GW energy $\Delta E_{\rm GW}$ (Eq. \eqref{EGWu}) are connected by the Mellin transform (Eq. \eqref{mellin}) of the function $\mathcal K(u)$, being defined in terms of the integrals 
\beq
I_{W_1}=\int_0^\infty du \mathcal K(u) \ln u
\eeq
and
\beq
I_{\Delta E}=\int_0^\infty du \mathcal K(u)\,,
\eeq
respectively.
Denoting by $f(u)=\mathcal K(u)$ and by $g(s)$ its Mellin transform, we then have that $I_{\Delta E}=g(1)$ and $I_{W_1}=\frac{dg(s)}{ds}\big|_{s=1}$.
Mellin transforms are well implemented in standard symbolic algebra manipulators.

At the Newtonian level, the function $\mathcal K(u)$ is expressed in terms of modified Bessel functions of the second kind. The typical term has the form
\beq
u^kK_\mu(u)K_\nu(u)\,,
\eeq
so that it is enough to compute the Mellin transform $g_{\rm KK}(s;\mu,\nu)$ of the function 
\beq
f_{\rm KK}(u;\mu,\nu)=K_\mu(u)K_\nu(u)\,,
\eeq
(see Eq. \eqref{gkk}) and its first derivative with respect to $s$, further using the property $\mathfrak M\{x^kf(x);s\}=g(s+k)$.

At higher PN orders also appear terms like
\beq
u^kK_\nu(u)\cos(u\sinh v)\,, \qquad
u^kK_\nu(u)\sin(u\sinh v)\,,
\eeq
to be integrated both  over $u$ and $v$.
Hence we also need the Mellin transform $g_{\rm Kcos}(s;v,\nu)$ and $g_{\rm Ksin}(s;v,\nu)$ of the functions
\bea
f_{\rm Kcos}(u;v,\nu)&=&K_\nu(u)\cos(u\sinh v)\,,\nonumber\\
f_{\rm Ksin}(u;v,\nu)&=&K_\nu(u)\sin(u\sinh v)\,,
\eea
and their first derivatives with respect to $s$.
Their Mellin transforms are given by
\bea
\label{gdefs}
g_{\rm Kcos}(s;\nu,v) &=&  
\frac{2^{s-2}}{\cosh^{s-\nu}v}\Gamma\left(\frac{s+\nu}{2}\right) \Gamma\left(\frac{s-\nu}{2}\right)\nonumber\\
&\times&
{}_{2}F_{1}\left(\frac{1-s-\nu}{2}, \frac{s-\nu}{2};\frac12;\tanh^2v\right)
\,,\nonumber\\
g_{\rm Ksin}(s;\nu,v)&=&
\frac{2^{s-1}\sinh v}{\cosh^{1+s+\nu}v}\Gamma\left(\frac{s+\nu+1}{2}\right) \Gamma\left(\frac{s-\nu+1}{2}\right)\nonumber\\
&\times&
{}_{2}F_{1}\left(\frac{2-s+\nu}{2}, \frac{s+\nu+1}{2};\frac32;\tanh^2v\right)
\,. \nonumber\\
\eea
For each of them ($i=$ KK, Kcos, Ksin) we need then
\beq
g_{i s}=\frac{\partial}{\partial s}g_i\,, \qquad 
g_{i \nu\nu  }=\frac{\partial^2}{\partial \nu^2} g_i\,, \qquad
g_{i s \nu\nu }=\frac{\partial^3}{\partial s \partial \nu^2} g_i\,,
\eeq
(for example, $g_{{\rm KK} s}=g_{{\rm KK} s}(s;\mu,\nu)$, etc.) and higher derivatives with respect to the order $\nu$ for increasing PN accuracy as well as level of expansion in the eccentricity parameter.
Explicit expressions can be obtained which generally involve HPLs, coming from the derivatives of the hypergeometric functions with respect to their parameters $(s,\nu)$ (see below).

\subsection{Results}

The function $\mathcal K(u)$ can be decomposed as in Eq. \eqref{eq_718} (here in a conveniently rescaled form)
\beq
\mathcal K(u)=\mathcal K_{\rm N}(u)+\frac{\eta^2}{\bar a_r}\mathcal K_{\rm 1PN}(u)+\frac{\eta^4}{\bar a_r^2}\mathcal K_{\rm 2PN}(u)\,,
\eeq
with 
\bea
\mathcal K_{\rm nPN}(u)&=&\frac{\nu^2}{e_r^2\bar a_r^2}\left[\tilde {\mathcal K}_{\rm nPN}^{\rm LO}(u)+\frac{\pi}{e_r}\tilde {\mathcal K}_{\rm nPN}^{\rm NLO}(u)\right.\nonumber\\
&+&\left.
\frac{1}{e_r^2}\tilde {\mathcal K}_{\rm nPN}^{\rm NNLO}(u)+\frac{\pi}{e_r^3}\tilde {\mathcal K}_{\rm nPN}^{\rm N^3LO}(u)\right]\,,
\eea
up to the 2PN order and to the N$^3$LO order in the large eccentricity.

At the Newtonial level we find
\begin{widetext}
\bea
\tilde {\mathcal K}_{\rm N}^{\rm LO}(u)&=&\frac{32}{5}u^2\left[
\left(\frac{1}{3}+u^2\right)K_0^2(u)+3u K_0(u)K_1(u)+(1+u^2)K_1^2(u)
\right]
\,,\nonumber\\
\tilde {\mathcal K}_{\rm N}^{\rm NLO}(u)&=&u\tilde {\mathcal K}_{\rm N}^{\rm LO}(u)
\,,\nonumber\\
\tilde {\mathcal K}_{\rm N}^{\rm NNLO}(u)&=&\frac{\pi^2}{2}u^2\tilde {\mathcal K}_{\rm N}^{\rm LO}(u)
-\frac{32}{5}u^2\left\{
(1+3u^2)K_0^2(u)+7u K_0(u)K_1(u)+(1+2u^2)K_1^2(u)\right.\nonumber\\
&&\left.
+u^2\left[\left(u^2+\frac13\right)K_0(u)+\frac32uK_1(u)\right]\frac{\partial^2 K_\nu(u)}{\partial \nu^2}\Bigg|_{\nu=0}
+u^2\left[\frac32uK_0(u)+(u^2+1)K_1(u)\right]\frac{\partial^2 K_\nu(u)}{\partial \nu^2}\Bigg|_{\nu=1}
\right\}
\,,\nonumber\\
\tilde {\mathcal K}_{\rm N}^{\rm N^3LO}(u)&=&u\tilde {\mathcal K}_{\rm N}^{\rm NNLO}(u)-\frac{\pi^2}{3}u^3\tilde {\mathcal K}_{\rm N}^{\rm LO}(u)
\,.
\eea
Using the same decomposition as above for $I_{\Delta E,\,\rm N}$ we find
\bea
\label{IDeltaEN}
I_{\Delta E,\,\rm N}^{\rm LO}&=&\frac{32}{15}[3 g_{\rm KK}(5;1,1)+3 g_{\rm KK}(3;1,1)+9  g_{\rm KK}(4;0,1)+3 g_{\rm KK}(5;0,0)+ g_{\rm KK}(3;0,0)]\nonumber\\
&=&\frac{37}{15}\pi^2 
\,,\nonumber\\
I_{\Delta E,\,\rm N}^{\rm NLO}&=&\frac{32}{15}[3 g_{\rm KK}(6;1,1)+3 g_{\rm KK}(4;1,1)+9  g_{\rm KK}(5;0,1)+3 g_{\rm KK}(6;0,0)+ g_{\rm KK}(4;0,0)]\nonumber\\
&=&\frac{1568}{45}
\,,\nonumber\\
I_{\Delta E,\,\rm N}^{\rm NNLO}&=&\frac{16}{15}[-9 g_{{\rm KK}\nu\nu}(6;1,0)-6 g_{{\rm KK}\nu\nu}(7;0,0)-2 g_{{\rm KK}\nu\nu}(5;0,0) ] \nonumber\\
&+&\frac{16}{15}[-9 g_{{\rm KK}\nu\nu}(6;0,1)-6 g_{{\rm KK}\nu\nu}(7;1,1)-6 g_{{\rm KK}\nu\nu}(5;1,1) ] \nonumber\\
&+& \frac{16}{15}\left[3\pi^2 \left( g_{\rm KK}(7;0,0)+ g_{\rm KK}(7;1,1)+ g_{\rm KK}(5;1,1)+3 g_{\rm KK}(6;0,1) +\frac13  g_{\rm KK}(5;0,0) \right)\right. \nonumber\\
&&\left. -12 g_{\rm KK}(5;1,1)-6  g_{\rm KK}(3;1,1)-42  g_{\rm KK}(4;0,1)-18  g_{\rm KK}(5;0,0)-6  g_{\rm KK}(3;0,0)  \right]\nonumber\\
&=&\frac{281}{10}\pi^2
\,,\nonumber\\
I_{\Delta E,\,\rm N}^{\rm N^3LO}&=&\frac{16}{45}[-27  g_{{\rm KK}\nu\nu}(7;1,0)-18  g_{{\rm KK}\nu\nu}(8;0,0)-6  g_{{\rm KK}\nu\nu}(6;0,0)]\nonumber\\
&+& \frac{16}{45}[-27  g_{{\rm KK}\nu\nu}(7;0,1)-18  g_{{\rm KK}\nu\nu}(8;1,1)-18  g_{{\rm KK}\nu\nu}(6;1,1)]\nonumber\\
&& +\frac{16}{45}\left[3\pi^2 \left( g_{\rm KK}(8;1,1)+ g_{\rm KK}(6;1,1)+3 g_{\rm KK}(7;0,1)+\frac13  g_{\rm KK}(6;0,0)\right)\right.\nonumber\\
&&\left.-18  g_{\rm KK}(4;1,1)-36  g_{\rm KK}(6;1,1)-126  g_{\rm KK}(5;0,1)-54  g_{\rm KK}(6;0,0)-18  g_{\rm KK}(4;0,0)
\right]\nonumber\\
&=&\frac{7808}{45}
\,,
\eea
where the values of the various Mellin transforms are listed in Table \ref{table_mellin1}.
The corresponding result for $I_{W_1,\,\rm N}$ is obtained simply by replacing each of them by its derivative with respect to the Mellin parameter, leading to 
\bea
\label{IW1N}
I_{W_1,\,\rm N}^{\rm LO}&=&
\left(\frac{40}{3}-\frac{74}{5}\ln(2)-\frac{74}{15}\gamma\right)\pi^2
\,,\nonumber\\
I_{W_1,\,\rm N}^{\rm NLO}&=&
\frac{4448}{135}+\frac{3136}{45}\ln(2)-\frac{3136}{45}\gamma
\,,\nonumber\\
I_{W_1,\,\rm N}^{\rm NNLO}&=&
\left(\frac{2479}{30}-\frac{843}{5}\ln(2)-\frac{281}{5}\gamma+\frac{2079}{20}\zeta(3)\right)\pi^2
\,,\nonumber\\
I_{W_1,\,\rm N}^{\rm N^3LO}&=&
-\frac{23936}{675}+\frac{15616}{45}\ln(2)-\frac{15616}{45}\gamma+\frac{88576}{225}\zeta(3)
\,.
\eea


\begin{table*}
\caption{\label{table_mellin1} List of Mellin transforms \eqref{gkk} used in Eqs. \eqref{IDeltaEN}--\eqref{IW1N}.
The function $g_{\rm KK}(s; \mu, \nu)$ and its derivative with respect to the Mellin parameter $s$ are both evaluated at $s=1+k$.}
\begin{ruledtabular}
\begin{tabular}{|l|l|l|l|l|l|l|}
$1+k$ & $\mu $ & $\nu$ & $ g_{\rm KK}$  & $  g_{{\rm KK} s}$ & $  g_{{\rm KK} \nu\nu}$ & $  g_{{\rm KK} s\nu\nu }$\\
\hline
3 & 0 & 0 & $\frac{\pi^2}{32}$ &$-\frac{\pi^2(6\ln(2)-5+2\gamma)}{64} $&$\frac{\pi ^2 }{64} \left(\pi ^2-8\right)$&
$-\frac{7 \pi ^2 \zeta (3)}{32}-\frac{\pi ^2}{16}+\frac{\gamma  \pi ^2}{8}+\frac{5 \pi ^4}{128}$\\
&&&&&& $-\frac{\gamma  \pi ^4}{64}+\frac{3}{8} \pi ^2 \ln (2)-\frac{3}{64} \pi ^4 \ln (2)$\\
3 & 0 & 1 & $\frac12$ &$\frac14(2\ln 2 -2\gamma -1)$&$\frac{(3+\pi^2)}{12}$&$-\frac{\zeta (3)}{2}-\frac{3}{8}-\frac{\gamma }{4}-\frac{\pi ^2}{24}$\\
&&&&&& $-\frac{\gamma  \pi ^2}{12}+\frac{\ln (2)}{4}+\frac{1}{12} \pi ^2 \ln (2)$\\
3 & 1 & 1 & $\frac{3\pi^2}{32}$ &$-\frac{\pi^2 (18\ln(2)-11+6\gamma)}{64}$&$\frac{\pi ^2 }{64} \left(3\pi ^2-8\right)$&$-\frac{21 \pi ^2 \zeta (3)}{32}+\frac{\pi ^2}{16}+\frac{\gamma  \pi ^2}{8}+\frac{11 \pi ^4}{128}$\\
&&&&&& $-\frac{3 \gamma  \pi ^4}{64}+\frac{3}{8} \pi ^2 \ln (2)-\frac{9}{64} \pi ^4 \ln (2)$\\
\hline
4 & 0 & 0 & $\frac{1}{3}$ & $\frac13 \ln(2)+\frac{1}{18}-\frac13 \gamma$ &$\frac{(\pi^2-6)}{18}$&$\frac{5}{18}+\frac{\gamma }{3}+\frac{\pi ^2}{108}-\frac{\gamma  \pi ^2}{18}$\\
&&&&&& $-\frac{\ln (2)}{3}+\frac{1}{18} \pi ^2 \ln (2)-\frac13\zeta(3)$\\
4 & 0 & 1 & $\frac{3\pi^2}{64}$ &$ -\frac{\pi^2 (18\ln(2)-17+6\gamma)}{128} $&$\frac{1}{384} \pi ^2 \left(9 \pi ^2-68\right)$&$-\frac{21 \pi ^2 \zeta (3)}{64}-\frac{77 \pi ^2}{576}+\frac{17 \gamma  \pi ^2}{96}$\\
&&&&&& $+\frac{17 \pi ^4}{256}-\frac{3 \gamma  \pi ^4}{128}+\frac{17}{32} \pi ^2 \ln (2)-\frac{9}{128} \pi
   ^4 \ln (2)$\\
4 & 1 & 1 & $\frac{2}{3}$ & $\frac23 \ln(2)-\frac{1}{18}-\frac23 \gamma $&$\frac{(2\pi^2-3)}{18}$&$-\frac{2 \zeta (3)}{3}+\frac{2}{9}+\frac{\gamma }{6}-\frac{\pi ^2}{108}$\\
&&&&&& $-\frac{\gamma  \pi ^2}{9}-\frac{\ln (2)}{6}+\frac{1}{9} \pi ^2 \ln (2)$\\
\hline
5 & 0 & 0 & $\frac{27\pi^2}{512}$ & $-\frac{27\pi^2(12\ln(2)-13+4\gamma) }{2048} $&$\frac{3\pi^2 (-80+9\pi^2)}{1024}$&$-\frac{83}{256}\pi^2-\frac{189}{512}\pi^2\zeta(3)+\frac{351}{4096}\pi^4+\frac{15}{64}\gamma\pi^2$\\
&&&&&&$-\frac{27}{1024}\pi^4\gamma+\frac{45}{64}\ln(2)\pi^2-\frac{81}{1024}\pi^4\ln(2)$\\
5 & 0 & 1 & $\frac{2}{3}$ & $\frac23 \ln(2)+\frac{5}{18}-\frac23 \gamma $&$\frac{(-21+4\pi^2)}{36}$&$-\frac{2 \zeta (3)}{3}+\frac{55}{144}+\frac{7 \gamma }{12}$\\
&&&&&&$+\frac{5 \pi ^2}{108}-\frac{\gamma  \pi ^2}{9}-\frac{7 \ln (2)}{12}+\frac{1}{9} \pi ^2 \ln (2)$\\
5 & 1 & 1 & $\frac{45\pi^2}{512}$ &$ -\frac{3\pi^2 (180\ln(2)-187+60\gamma)}{2048}$&$\frac{\pi^2(-368+45\pi^2)}{1024}$&$\frac{23}{64}\gamma\pi^2+\frac{561}{4096}\pi^4-\frac{315}{512}\pi^2\zeta(3)+\frac{69}{64}\ln(2)\pi^2$\\
&&&&&&$-\frac{105}{256}\pi^2-\frac{45}{1024}\pi^4\gamma-\frac{135}{1024}\pi^4\ln(2)$\\
\hline
6 & 0 & 0 & $\frac{16}{15}$ &$ \frac{16}{15}\ln(2)+\frac{172}{225}-\frac{16}{15}\gamma $&$-\frac43 +\frac{8}{45}\pi^2 $&$\frac{11}{45}-\frac{16}{15}\zeta(3)+\frac{86}{675}\pi^2+\frac43 \gamma-\frac{8}{45}\gamma\pi^2-\frac43 \ln(2)$\\
&&&&&&$+\frac{8}{45}\ln(2)\pi^2$\\
6 & 0 & 1 & $\frac{135\pi^2}{1024}$ & $-\frac{27\pi^2(60\ln(2)-69+20\gamma)}{4096}  $&$ \frac{3\pi^2(-1964+225\pi^2) }{10240}$&$\frac{1473}{2560}\gamma\pi^2+\frac{1863}{8192}\pi^4-\frac{945}{1024}\pi^2\zeta(3)-\frac{45853}{51200}\pi^2$\\
&&&&&&$+\frac{4419}{2560}\ln(2)\pi^2-\frac{135}{2048}\pi^4\gamma-\frac{405}{2048}\pi^4\ln(2)$\\
6 & 1 & 0 & $\frac{135\pi^2}{1024}$&$-\frac{27\pi^2(60\ln(2)-69+20\gamma)}{4096}  $ &$ \frac{3\pi^2(-2036+225\pi^2)}{10240}$&$-\frac{49147}{51200}\pi^2+\frac{1527}{2560}\gamma\pi^2+\frac{1863}{8192}\pi^4-\frac{945}{1024}\pi^2\zeta(3)$\\
&&&&&& $+\frac{4581}{2560}\ln(2)\pi^2-\frac{135}{2048}\pi^4\gamma-\frac{405}{2048}\pi^4\ln(2)$\\ 
6 & 1 & 1 & $\frac{8}{5}$ & $ \frac85 \ln(2)+\frac{76}{75}-\frac{8}{5}\gamma $&$-\frac53 +\frac{4}{15}\pi^2$&$\frac53 \gamma+\frac{38}{225}\pi^2-\frac{8}{5}\zeta(3)-\frac53 \ln(2)+\frac{53}{90}-\frac{4}{15}\gamma\pi^2$\\
&&&&&& $+\frac{4}{15}\ln(2)\pi^2$\\
\hline
7 & 0 & 0 & $\frac{1125\pi^2}{4096}$ &$-\frac{75\pi^2(180\ln(2)-221+60\gamma) }{16384} $&$\frac{5\pi^2(-2072+225\pi^2)}{8192} $&$-\frac{29023}{12288}\pi^2-\frac{7875}{4096}\pi^2\zeta(3)+\frac{16575}{32768}\pi^4+\frac{1295}{1024}\gamma\pi^2$\\
&&&&&& $-\frac{1125}{8192}\pi^4\gamma+\frac{3885}{1024}\ln(2)\pi^2-\frac{3375}{8192}\pi^4\ln(2)$\\
7 & 0 & 1 & $\frac{16}{5}$ & $\frac{16}{5}\ln(2)+\frac{212}{75}-\frac{16}{5}\gamma $&$-\frac{172}{45}+\frac{8}{15}\pi^2$ &$\frac{172}{45}\gamma+\frac{106}{225}\pi^2-\frac{16}{5}\zeta(3)-\frac{172}{45}\ln(2)+\frac{37}{225}$\\
&&&&&&$-\frac{8}{15}\gamma \pi^2+\frac{8}{15}\ln(2)\pi^2$ \\
7 & 1 & 0 & $\frac{16}{5}$ &$\frac{16}{5}\ln(2)+\frac{212}{75}-\frac{16}{5}\gamma $& $-\frac{188}{45}+\frac{8}{15}\pi^2$ &$\frac{188}{45}\gamma+\frac{106}{225}\pi^2-\frac{16}{5}\zeta(3)-\frac{188}{45}\ln(2)-\frac{8}{15}\gamma \pi^2$\\
&&&&&&$+\frac{8}{15}\ln(2)\pi^2-\frac{7}{225}$\\
7 & 1 & 1 & $\frac{1575\pi^2}{4096}$ &$-\frac{15\pi^2(1260\ln(2)-1523+420\gamma) }{16384} $&$\frac{\pi^2(-14072+1575\pi^2)}{8192}$&$\frac{1759}{1024}\gamma\pi^2+\frac{22845}{32768}\pi^4-\frac{11025}{4096}\pi^2\zeta(3)+\frac{5277}{1024}\ln(2)\pi^2$\\
&&&&&& $-\frac{37283}{12288}\pi^2-\frac{1575}{8192}\pi^4\gamma-\frac{4725}{8192}\pi^4\ln(2)$\\
\hline
8 & 0 & 0 & $\frac{288}{35}$ & $\frac{288}{35}\ln(2)+\frac{10824}{1225}-\frac{288}{35}\gamma $&$-\frac{56}{5}+\frac{48}{35}\pi^2$&$-\frac{1294}{525}+\frac{1804}{1225}\pi^2+\frac{56}{5}\gamma-\frac{48}{35}\gamma \pi^2-\frac{56}{5}\ln(2)$\\
&&&&&& $+\frac{48}{35}\ln(2)\pi^2-\frac{288}{35}\zeta(3)$\\
8 & 0 & 1 & $\frac{7875\pi^2}{8192}$ &$\frac{120525 \pi ^2}{32768}-\frac{7875 \gamma  \pi ^2}{8192}-\frac{23625 \pi ^2
   \ln (2)}{8192}$&$ \frac{5 \pi ^2 \left(11025 \pi ^2-100628\right)}{114688} $&$-\frac{55125 \pi ^2 \zeta (3)}{8192}-\frac{21098123 \pi ^2}{2408448}+\frac{125785 \gamma  \pi ^2}{28672}+\frac{120525 \pi ^4}{65536}$\\
&&&&&& $-\frac{7875 \gamma  \pi ^4}{16384}+\frac{377355 \pi
   ^2 \ln (2)}{28672}-\frac{23625 \pi ^4 \ln (2)}{16384}$\\
8&1&0 &$\frac{7875\pi^2}{8192}$ &$\frac{120525 \pi ^2}{32768}-\frac{7875 \gamma  \pi ^2}{8192}-\frac{23625 \pi ^2
   \ln (2)}{8192}$&$\frac{5 \pi ^2 \left(11025 \pi ^2-102428\right)}{114688}$&$-\frac{55125 \pi ^2 \zeta (3)}{8192}-\frac{21767273 \pi ^2}{2408448}+\frac{128035 \gamma  \pi ^2}{28672}$\\
&&&&&& $+\frac{120525 \pi ^4}{65536}-\frac{7875 \gamma  \pi ^4}{16384}+\frac{384105 \pi
   ^2 \ln (2)}{28672}-\frac{23625 \pi ^4 \ln (2)}{16384}$\\
8 & 1 & 1 & $\frac{384}{35}$ & $\frac{384}{35}\ln(2)+\frac{13872}{1225}-\frac{384}{35}\gamma $&$-\frac{208}{15}+\frac{64}{35}\pi^2$&$\frac{208}{15}\gamma+\frac{2312}{1225}\pi^2-\frac{384}{35}\zeta(3)-\frac{208}{15}\ln(2)-\frac{2992}{1575}$\\
&&&&&& $-\frac{64}{35}\gamma \pi^2+\frac{64}{35}\ln(2)\pi^2$\\
\hline
\end{tabular}
\end{ruledtabular}
\end{table*}

Starting from the 1PN level, the Fourier transform of the multipolar moments can be explicitly done only partly, so that the resulting function $\mathcal K(u)$ is not fully determined in closed form. 
Consider, for instance, the NLO term
\bea
\tilde {\mathcal K}_{\rm 1PN}^{\rm NLO}(u)&=&\frac{16}{21}u^3\left[
\left(u^4-46 u^2-\frac{141}{5}\right)K_0(u)^2+\frac{122}{5}u\left(u^2-\frac{653}{122}\right)K_0(u)K_1(u)+\left(u^4-\frac{333 u^2}{10}-\frac{39}{5}\right)K_1(u)^2
\right]\nonumber\\
&&
-\frac{48}{5\pi}u^4\int_{-\infty}^{\infty} dv\, {\rm arctan}\left(\tanh\frac{v}{2}\right)\left[
\sinh 2v(K_0(u)+2uK_1(u))\cos(u\sinh v)\right.\nonumber\\
&&\left.
+\frac12(\cosh 3v-5\cosh v)(uK_0(u)+K_1(u))\sin(u\sinh v)
\right]\nonumber\\
&&
-\frac{64}{21}u^3\left[
\left(u^4-\frac{21 u^2}{20}-\frac{3}{4}\right)K_0(u)^2-\frac{6}{5}u\left(u^2+\frac{95}{24}\right)K_0(u)K_1(u)+\left(u^4-\frac{23 u^2}{20}-\frac{21}{20}\right)K_1(u)^2
\right]\nu
\,.\nonumber\\
\eea
It is convenient taking the Mellin transform first (i.e., integrating over $u$), and then integrating over $v$.
We find
\bea
I_{\Delta E,\,\rm 1PN}^{\rm NLO}
&=&-\frac{888}{35} g_{\rm KK}(6;1,1)-\frac{208}{35} g_{\rm KK}(4;1,1)+\frac{16}{21} g_{\rm KK}(8;1,1)+\frac{1952}{105} g_{\rm KK}(7;0,1)-\frac{10448}{105} g_{\rm KK}(5;0,1)\nonumber\\
&+&\frac{16}{21} g_{\rm KK}(8;0,0)-\frac{752}{35} g_{\rm KK}(4;0,0)-\frac{736}{21} g_{\rm KK}(6;0,0)\nonumber\\
&+&\frac{1}{\pi}\int dv {\rm arctan}\left(\tanh\left(\frac{v}{2}\right)\right)\times  \nonumber\\
&-&\left[\frac{48}{5}\sinh(2 v) (2 g_{\rm Kcos}(6;1,v)+g_{\rm Kcos}(5;0,v))  
+\frac{96}{5} \cosh(v) (\cosh(v)^2-2) (g_{\rm Ksin}(5;1,v)+g_{\rm Ksin}(6;0,v)) \right]\nonumber\\
&&
+\left[
-\frac{64}{21} g_{\rm KK}(8;1,1)+\frac{368}{105} g_{\rm KK}(6;1,1)+\frac{16}{5} g_{\rm KK}(4;1,1)+\frac{128}{35} g_{\rm KK}(7;0,1)+\frac{304}{21} g_{\rm KK}(5;0,1)\right.\nonumber\\
&&\left.-\frac{64}{21} g_{\rm KK}(8;0,0)+\frac{16}{5} g_{\rm KK}(6;0,0)+\frac{16}{7} g_{\rm KK}(4;0,0)\right]\nu
\nonumber\\
&=&-\frac{25616}{315}
+\int dv\, {\rm arctan}\left(\tanh\left(\frac{v}{2}\right)\right)\frac{\sinh v}{\cosh^4 v}\left(-\frac{4032}{5}+\frac{2448}{\cosh^2 v}\right)
-\frac{1136}{45}\nu
\nonumber\\
&=&
\frac{944}{1575}-\frac{1136}{45}\nu
\,,
\eea
where we have used 
\bea
g_{\rm Kcos}(5;0,v)&=&\frac{3\pi}{2\cosh^9v} (8 \cosh^4v-40 \cosh^2v+35)
\,,\nonumber\\
g_{\rm Kcos}(6;1,v)&=&\frac{45\pi}{2\cosh^{11}v} (8 \cosh^4v-28 \cosh^2v+21)
\,,\nonumber\\
g_{\rm Ksin}(5;1,v)&=&-\frac{15\pi\sinh v}{2\cosh^9v} (4 \cosh^2v-7) 
\,,\nonumber\\
g_{\rm Ksin}(6;0,v)&=&\frac{15\pi\sinh v}{2\cosh^{11}v}  (8 \cosh^4v-56 \cosh^2v+63) 
\,.
\eea
The corresponding result for $I_{W_1,\,\rm 1PN}^{\rm NLO}$ is
\bea
I_{W_1,\,\rm 1PN}^{\rm NLO}&=&
-\frac{3536}{135}-\frac{51232}{315}\ln(2)+\frac{51232}{315}\gamma\nonumber\\
&&
+\int dv\, {\rm arctan}\left(\tanh\left(\frac{v}{2}\right)\right)\frac{\sinh v}{\cosh^2 v}\left[
-\frac{576}{5}+\left(\frac{8064}{5}\ln(2)+\frac{8064}{5}\gamma+\frac{16128}{5}\ln(\cosh v)-\frac{34464}{5}\right)\frac1{\cosh^2v}\right.\nonumber\\
&&\left.
+\left(-9792\ln(\cosh v)+\frac{86592}{5}-4896\ln(2)-4896\gamma\right)\frac1{\cosh^4v}
\right]\nonumber\\
&&
+\left(-\frac{77744}{945}-\frac{2272}{45}\ln(2)+\frac{2272}{45}\gamma\right)\nu\nonumber\\
&=&
-\frac{56144}{3375}+\frac{1888}{1575}\ln(2)-\frac{1888}{1575}\gamma
+\left(-\frac{77744}{945}-\frac{2272}{45}\ln(2)+\frac{2272}{45}\gamma\right)\nu\,,
\eea
where we have used 
\bea
g_{{\rm Kcos}s}(5;0,v)&=&-\frac{12\pi}{\cosh^9v} \left[
\left(2\cosh^4v-10\cosh^2v+\frac{35}{4}\right)\ln(\cosh v)+\left(\gamma-\frac{25}{6}+\ln(2)\right)\cosh^4v\right.\nonumber\\
&&\left.
+\left(\frac{107}{6}-5\ln(2)-5\gamma\right)\cosh^2v+\frac{35}{8}\ln(2)-\frac{44}{3}+\frac{35}{8}\gamma
\right]
\,,\nonumber\\
g_{{\rm Kcos}s}(6;1,v)&=&-\frac{180\pi}{\cosh^{11}v} \left[
\left(2\cosh^4v-7\cosh^2v+\frac{21}{4}\right)\ln(\cosh v)+\frac{1}{15}\cosh^6v+\left(\gamma+\ln(2)-\frac{127}{30}\right)\cosh^4v\right.\nonumber\\
&&\left.
+\left(-\frac{7}{2}\ln(2)+\frac{1583}{120}-\frac{7}{2}\gamma\right)\cosh^2v+\frac{21}{8}\ln(2)+\frac{21}{8}\gamma-\frac{563}{60}
\right]
\,,\nonumber\\
g_{{\rm Ksin}s}(5;1,v)&=&\frac{\pi \sinh v}{2\cosh^9v}\left[
(120\cosh^2v-210)\ln(\cosh v)+6\cosh^4v\right.\nonumber\\
&&\left.
+(60\gamma-229+60\ln(2))\cosh^2v-105\gamma+352-105\ln(2)
\right]
\,,\nonumber\\
g_{{\rm Ksin}s}(6;0,v)&=&-\frac{60\pi \sinh v}{\cosh^{11}v} \left[
\left(\frac{63}{4}+2\cosh^4v-14\cosh^2v\right)\ln(\cosh v)+\left(\gamma-\frac{137}{30}+\ln(2)\right)\cosh^4v\right.\nonumber\\
&&\left.
+\left(\frac{809}{30}-7\gamma-7\ln(2)\right)\cosh^2v+\frac{63}{8}\ln(2)-\frac{563}{20}+\frac{63}{8}\gamma
\right]
\,.
\eea
At the NNLO the derivatives of the hypergeometric functions entering the Mellin transforms \eqref{gdefs} also generates HPLs of weight 2.
Consider, for instance, the Mellin transform $g_{\rm Kcos}(6;0,v)$ and its derivative $g_{{\rm Kcos}s}(6;0,v)$.
We find
\beq
g_{\rm Kcos}(6;0,v)=\left(-\frac{120}{\cosh^7v}+\frac{840}{\cosh^9v}-\frac{945}{\cosh^{11}v}\right)v\sinh v+\frac{274}{\cosh^6v}-\frac{1155}{\cosh^8v}+\frac{945}{\cosh^{10}v}\,,
\eeq
and 
\bea
g_{{\rm Kcos}s}(6;0,v)&=&
-g_{\rm Kcos}(6;0,v)\left(\ln(\cosh v)-\ln(2)+\gamma-\frac32\right)\nonumber\\
&&
+\frac{64}{\cosh^6v}\frac{\partial}{\partial s}\,{}_{2}F_{1}\left(\frac{s}{2}, \frac{1-s}{2}; \frac12;\tanh(v)^2\right) \bigg|_{s=6} 
\eea
respectively.
The latter term can be computed, e.g., by using the tool \texttt{HypExp2} \cite{Huber:2007dx}, which allows for Taylor-expanding hypergeometric functions around their parameters. It reads
\bea \label{dhypergeom}
\frac{\partial}{\partial s}\,{}_{2}F_{1}\left(\frac{s}{2}, \frac{1-s}{2}; \frac12;\tanh(v)^2\right) \bigg|_{s=6}&=&
\left[\left(-\frac{15}{8\cosh v}+\frac{105}{8\cosh^3v}-\frac{945}{64\cosh^5v}\right) v\sinh v\right.\nonumber\\
&&\left.
+\frac{945}{64 \cosh^4v}-\frac{1155}{64\cosh^2v}+\frac{137}{32}\right]\ln(\cosh v) \nonumber\\
&&+\left(\frac{945}{128 \cosh^5v} -\frac{105}{16\cosh^3v}  +\frac{15}{16\cosh v}\right)|\sinh v|H_{-,+}(|\tanh v|)\nonumber\\
&&+\left(\frac{247}{8\cosh^3v}-\frac{3921}{128 \cosh^5v}-\frac{23}{4\cosh v}\right) v\sinh v\nonumber\\
&&
+\frac{141}{128 \cosh^4v}+\frac{39}{64}-\frac{219}{128\cosh^2v}\,,
\eea
\end{widetext}
where 
\bea
H_{-,+}(|\tanh v|)&=& 
2\ln (2\cosh v)|v|+{\rm Li}_2\left(\frac12-\frac{|\sinh v|}{2\cosh v}\right)\nonumber\\
&&
-{\rm Li}_2\left(\frac12+\frac{|\sinh v|}{2\cosh v}\right) \,,
\eea
is an HPL with weights $\pm$, which can be in turn converted into HPLs with integer weights according to the rule
\beq
H_{-,+}(x)= - H_{-1,-1}(x) - H_{-1,1}(x)+ H_{1,-1}(x)+H_{1,1}(x)\,.
\eeq

Going to the 2PN level we get more involved expressions, but with the same structure (further including terms containing derivatives of the Bessel functions with respect to the order up to the fourth at N$^3$LO as well as HPLs of increasing weight).

\section{Summary of final results for the integrated nonlocal action}

We recap below our final results for the integrated nonlocal action $W^{\rm tail, h}$ up to the N$^3$LO order in the large eccentricity expansion, showing also equivalent forms corresponding to different choices of orbital parameters used as independent variables, i.e., either $(\bar a_r, e_r)$ or $(\bar E, j)$, which are related by Eq. \eqref{arer}.

\subsection{First-order-tail part}

The 2PN-accurate values of the two contributions to the first-order tail $W^{\rm tail, h}=W_1^{\rm tail, h} +W_2^{\rm tail, h}$, i.e.,
\bea
\label{W12arer}
W_{1,2}^{\rm tail, h}&=& W_{1,2}^{\rm tail, h \, LO} + W_{1,2}^{\rm tail, h \, NLO}  \nonumber\\
&+&
W_{1,2}^{\rm tail, h \, NNLO}+W_{1,2}^{\rm tail, h \, N^3LO}+ O(e_r^{-7})
\,,\nonumber\\
\eea
are listed in Tables \ref{W1_table} and \ref{W2_table}.
It is easily seen that the intermediate scale $s$ cancels between the two contributions.

Re-expressing $\bar a_r$ and $e_r$ in terms of $\bar E$ and $j$ we get
\beq
\label{WEj}
W^{\rm tail, h}=\frac{{\mathcal W}_3}{j^3}+\frac{{\mathcal W}_4}{j^4}
+\frac{{\mathcal W}_5}{j^5}+\frac{{\mathcal W}_6}{j^6}+O(j^{-7})\,.
\eeq
with coefficients ${\mathcal W}_k$, $k=3,4,5,6$ listed in Table \ref{mathcal_Wk_table}.


\begin{table*}
\caption{\label{W1_table} Expressions for the various coefficients $W_1^{\rm tail,h\,nLO}$ of the large-$e_r$ expansion \eqref {W12arer} of the first-order-tail $W_1^{\rm tail,h}$.
}
\begin{ruledtabular}
\begin{tabular}{|l|l|}
Coefficient & Expression   \\
\hline
$W_1^{\rm tail,h\,LO}$ & $\frac{2}{15} \frac{\pi M\nu^2}{e_r^3\bar a_r^{7/2}}H_{\rm tot}\left\{
100 + 37 \ln \left(\frac{s}{4e_r\bar a_r^{3/2}}\right)
+\left[\frac{685}{4}-\frac{1017}{14}\nu+\left(\frac{3429}{56}-\frac{37}{2}\nu\right)\ln \left(\frac{s}{4e_r\bar a_r^{3/2}}\right)\right] \frac{\eta^2}{\bar a_r}\right.$\\
&$\left.
+\left[
\frac{3656939}{8064}-\frac{18181}{72}\nu+\frac{235453}{4032}\nu^2
+\left(\frac{114101}{672}-\frac{7055}{112}\nu+\frac{111}{8}\nu^2\right)\ln \left(\frac{s}{4e_r\bar a_r^{3/2}}\right)
\right] \frac{\eta^4}{\bar a_r^2}
\right\}$   \\
$W_1^{\rm tail,h\,NLO}$ & $\frac{2}{15} \frac{M\nu^2}{e_r^4\bar a_r^{7/2}}H_{\rm tot}\left\{
\frac{2224}{9} + \frac{1568}{3} \ln \left(\frac{4s}{e_r\bar a_r^{3/2}}\right)
+\left[-\frac{28072}{225}-\frac{38872}{63}\nu+\left(\frac{944}{105}-\frac{1136}{3}\nu\right)\ln \left(\frac{4s}{e_r\bar a_r^{3/2}}\right)\right] \frac{\eta^2}{\bar a_r}\right.$\\
&$\left.
+\left[
-\frac{67489874}{77175}-\frac{3115726}{3675}\nu+\frac{165086}{315}\nu^2
+\left(\frac{419036}{735}-\frac{3244}{7}\nu+\frac{764}{3}\nu^2\right)\ln \left(\frac{4s}{e_r\bar a_r^{3/2}}\right)
\right] \frac{\eta^4}{\bar a_r^2}
\right\}$\\
$W_1^{\rm tail,h\,NNLO}$ & $\frac{2}{15} \frac{\pi M\nu^2}{e_r^5\bar a_r^{7/2}}H_{\rm tot}\left\{
\frac{2479}{4}+\frac{6237}{8}\zeta(3) + \frac{843}{2} \ln \left(\frac{s}{4e_r\bar a_r^{3/2}}\right)\right.$\\
&$
+\left[\frac{112309}{224}+\frac{27648}{7}\ln(2)-\frac{299511}{64}\zeta(3)+\left(-\frac{7332}{7}-918\zeta(3)\right)\nu\right.$\\
&$\left.
+\left(-\frac{66999}{112}-\frac{1827}{4}\nu\right)\ln \left(\frac{s}{4e_r\bar a_r^{3/2}}\right)\right] \frac{\eta^2}{\bar a_r}$\\
&$
+\left[
-\frac{26903663}{16128}-\frac{59760}{7}\ln(2)+\frac{571467}{128}\zeta(3)
+\left(\frac{2338541}{2688}-\frac{20224}{7}\ln(2)+\frac{918657}{256}\zeta(3)\right)\nu\right.$\\
&$\left.\left.
+\left(\frac{321719}{384}+\frac{35613}{64}\zeta(3)\right)\nu^2
+\left(-\frac{442237}{1344}+\frac{28735}{32}\nu+\frac{4497}{16}\nu^2\right)\ln \left(\frac{s}{4e_r\bar a_r^{3/2}}\right)
\right] \frac{\eta^4}{\bar a_r^2}
\right\}\,, $\\
$W_1^{\rm tail,h\, N^3LO}$ &$\frac{2}{15}\frac{M\nu^2}{e_r^6 \bar a_r^{7/2}}H_{\rm tot}\left\{
-\frac{11968}{45}+\frac{44288}{15}\zeta(3)
+\frac{7808}{3}\ln \left(\frac{4s}{\bar a_r^{3/2}e_r}\right)\right.$\\
&$
+\left[
-\frac{753568}{1575}-\frac{505984}{35}\zeta(3)+\frac{621}{8}\pi^4
+\left(-\frac{87136}{45}-\frac{577408}{105}\zeta(3)\right)\nu\right.$\\
&$\left.\left.
+\left(-\frac{763712}{105}-\frac{11584}{3}\nu\right)\ln \left(\frac{4s}{\bar a_r^{3/2}e_r}\right)
\right]\frac{\eta^2}{\bar a_r}\right\}$\\
&$
+\left[
\frac{261560660008}{22920975}+\frac{5617672}{3465}\zeta(3)+\frac{8424}{5}\pi^2-\frac{367551}{896}\pi^4\right.$\\
&$
+\left(\frac{954967984}{72765}+\frac{11922016}{495}\zeta(3)-\frac{26865}{224}\pi^4\right)\nu
+\left(\frac{20142088}{6615}+\frac{7589024}{2205}\zeta(3)\right)\nu^2$\\
&$\left.\left.
+\left(-\frac{119081008}{19845}+\frac{7488736}{315}\nu+2384\nu^2\right)\ln \left(\frac{4s}{\bar a_r^{3/2}e_r}\right)
\right]\frac{\eta^4}{\bar a_r^2}\right\}$\\
\end{tabular}
\end{ruledtabular}
\end{table*}


\begin{table*}
\caption{\label{W2_table} Expressions for the various coefficients $W_2^{\rm tail,h\,nLO}$ of the large-$e_r$ expansion \eqref {W12arer} of the first-order-tail $W_2^{\rm tail,h}$.
}
\begin{ruledtabular}
\begin{tabular}{|l|l|}
Coefficient & Expression   \\
\hline
$W_2^{\rm tail,h\, LO}$ &
$ \frac{2}{15} \frac{\pi M\nu^2}{e_r^3\bar a_r^{7/2}}H_{\rm tot}\left\{
-\frac{85}{4}-37\ln \left(\frac{s}{2e_r\bar a_r}\right)
+\left[
-\frac{9679}{224}+\frac{981}{56}\nu+\left(-\frac{3429}{56}+\frac{37}{2}\nu\right)\ln \left(\frac{s}{2e_r\bar a_r}\right)
\right] \frac{\eta^2}{\bar a_r}\right.$\\
&$\left.
+\left[
-\frac{1830565}{16128}+\frac{54899}{1152}\nu-\frac{29969}{4032}\nu^2
+\left(-\frac{114101}{672}+\frac{7055}{112}\nu-\frac{111}{8}\nu^2\right)\ln \left(\frac{s}{2e_r\bar a_r}\right)
\right] \frac{\eta^4}{\bar a_r^2}
\right\}$\\
$W_2^{\rm tail,h\, NLO}$
&$ \frac{2}{15} \frac{M\nu^2}{e_r^4\bar a_r^{7/2}}H_{\rm tot}\left\{
\frac{2768}{9}-\frac{1568}{3}\ln \left(\frac{2s}{e_r\bar a_r}\right)
+\left[
-\frac{64904}{225}-\frac{5992}{45}\nu+\left(-\frac{944}{105}+\frac{1136}{3}\nu\right)\ln \left(\frac{2s}{e_r\bar a_r}\right)
\right] \frac{\eta^2}{\bar a_r}\right.$\\
&$\left.
+\left[
-\frac{2925494}{77175}-\frac{542014}{2205}\nu+\frac{145498}{735}\nu^2
+\left(-\frac{419036}{735}+\frac{3244}{7}\nu-\frac{764}{3}\nu^2\right)\ln \left(\frac{2s}{e_r\bar a_r}\right)
\right] \frac{\eta^4}{\bar a_r^2}
\right\}$\\
$W_2^{\rm tail,h\, NNLO}$
&$ \frac{2}{15} \frac{\pi M \nu^2}{e_r^5\bar a_r^{7/2}}H_{\rm tot}\left\{
-\frac{3419}{8}-\frac{843}{2}\ln \left(\frac{s}{2e_r\bar a_r}\right)\right.$\\
&$
+\left[
\frac{103645}{448}+\frac{56559}{112}\nu+\left(\frac{66999}{112}+\frac{1827}{4}\nu\right)\ln \left(\frac{s}{2e_r\bar a_r}\right)
\right] \frac{\eta^2}{\bar a_r}$\\
&$\left.
+\left[
\frac{2467109}{13824}-\frac{3706175}{5376}\nu-\frac{1577635}{8064}\nu^2
+\left(\frac{442237}{1344}-\frac{28735}{32}\nu-\frac{4497}{16}\nu^2\right)\ln \left(\frac{s}{2e_r\bar a_r}\right)
\right] \frac{\eta^4}{\bar a_r^2}
\right\}$\\
$W_2^{\rm tail,h\, N^3LO}$&$\frac{2}{15}\frac{M\nu^2}{e_r^6 \bar a_r^{7/2}}H_{\rm tot}\left\{  
\frac{4384}{9}-\frac{7808}{3} \ln \left(\frac{2s}{e_r\bar a_r}\right)\right.$\\
&$
+\left[-\frac{7182736}{1575}-\frac{210928}{315}\nu
+\left(\frac{763712}{105}+\frac{11584}{3}\nu\right)\ln \left(\frac{2s}{e_r\bar a_r}\right)
\right] \frac{\eta^2}{\bar a_r}$\\
&$
+\left[-\frac{2135067428}{2083725}+\frac{392856316}{33075}\nu+\frac{1895524}{1323}\nu^2\right. $\\
&$\left.\left.
+\left(\frac{119081008}{19845}-\frac{7488736}{315}\nu-2384\nu^2\right)\ln \left(\frac{2s}{e_r\bar a_r}\right)
\right]\frac{\eta^4}{\bar a_r^2}\right\}$\\
\end{tabular}
\end{ruledtabular}
\end{table*}


\begin{table*}
\caption{\label{mathcal_Wk_table}  Coefficients ${\mathcal W}_n$ entering the large-$j$ expansion \eqref{WEj} of the first-order-tail $W^{\rm tail, h}=W_1^{\rm tail,h}+W_2^{\rm tail,h}$.
}
\begin{ruledtabular}
\begin{tabular}{|l|l|}
Coefficient & Expression   \\
\hline
${\mathcal W}_{3}$ &$\frac{\pi(2\bar E)^2}{15}M^2\nu^2\left\{
\frac{315}{2}+37\ln\left(\frac{\bar E}{2}\right)
+\left[\frac{12609}{112}-\frac{719}{4}\nu +\left(\frac{2393}{56}-37\nu\right)\ln\left(\frac{\bar E}{2}\right)\right](2\bar E)\eta^2\right.$\\
&$\left.
+\left[\frac{189347}{896}-\frac{295831}{2016}\nu+\frac{10603}{63}\nu^2
+\left(\frac{745}{12}-\frac{7179}{224}\nu+\frac{481}{16}\nu^2\right)\ln\left(\frac{\bar E}{2}\right)\right](2\bar E)^2\eta^4
\right\}$\\
${\mathcal W}_{4}$&$\frac{(2\bar E)^{3/2}}{15}M^2\nu^2\left\{
\frac{3328}{3}+\frac{1568}{3}\ln(8\bar E)
+\left[-\frac{16184}{75}-\frac{274856}{105}\nu+\left(\frac{25468}{35}-\frac{2900}{3}\nu\right)\ln(8\bar E)\right](2\bar E)\eta^2\right.$\\
&$\left.
+\left[-\frac{28646896}{15435}-\frac{931192}{1575}\nu+\frac{7860556}{2205}\nu^2
+\left(\frac{1894327}{2940}-\frac{67069}{70}\nu+\frac{4353}{4}\nu^2\right)\ln(8\bar E)\right](2\bar E)^2\eta^4
\right\}$\\
${\mathcal W}_{5}$& $\frac{\pi(2\bar E)}{15}M^2\nu^2\left\{
\frac{297}{2}+\frac{6237}{4}\zeta(3)+366\ln\left(\frac{\bar E}{2}\right)\right.$\\
&$
+\left[\frac{65547}{28}+\frac{55296}{7}\ln(2)-\frac{137349}{32}\zeta(3)+\left(-\frac{10593}{8}-\frac{73035}{16}\zeta(3)\right)\nu
+\left(\frac{46617}{56}-1125\nu\right)\ln\left(\frac{\bar E}{2}\right)\right](2\bar E)\eta^2$\\
&$
+\left[-\frac{388793}{6048}-\frac{36576}{7}\ln(2)-\frac{13845}{32}\zeta(3)
+\left(-\frac{7540333}{1344}-\frac{123392}{7}\ln(2)+\frac{1390239}{128}\zeta(3)\right)\nu\right.$\\
&$\left.\left.
+\left(\frac{218503}{72}+\frac{223533}{32}\zeta(3)\right)\nu^2
+\left(\frac{14975}{84}-\frac{24671}{16}\nu+\frac{6987}{4}\nu^2\right)\ln\left(\frac{\bar E}{2}\right)\right](2\bar E)^2\eta^4
\right\}\,,$\\
${\mathcal W}_{6}$&$\frac{(2\bar E)^{1/2}}{15}M^2\nu^2\left\{
-\frac{79936}{45}+\frac{88576}{15}\zeta(3)+\frac{4672}{3}\ln(8\bar E)\right.$\\
&$
+\left[-\frac{145784}{21}+\frac{28352}{21}\zeta(3)+\frac{621}{4}\pi^4+\left(\frac{56152}{315}-\frac{2627392}{105}\zeta(3)\right)\nu
+\left(\frac{146200}{21}-7848\nu\right)\ln(8\bar E)\right](2\bar E)\eta^2$\\
&$
+\left[-\frac{19308599}{2970}-\frac{19317988}{495}\zeta(3)+\frac{16848}{5}\pi^2-\frac{132813}{448}\pi^4
+\left(\frac{1528632323}{51975}+\frac{28349672}{3465}\zeta(3)-\frac{127629}{224}\pi^4\right)\nu\right.$\\
&$\left.\left.
+\left(\frac{9153757}{882}+\frac{111368548}{2205}\zeta(3)\right)\nu^2
+\left(-\frac{75949}{1134}-\frac{5855873}{315}\nu+\frac{33581}{2}\nu^2\right)\ln(8\bar E)\right](2\bar E)^2\eta^4
\right\}$\\
\end{tabular}
\end{ruledtabular}
\end{table*}


The $f$-induced additional contribution reads
\bea
\label{Wfmenharer}
W^{\rm f-h}&=&\frac{M\nu^3}{\bar a_r^{9/2}}H_{\rm tot}\eta^2\left[
\frac{\pi}{e_r^3}\mathcal W^{\rm f-h\,LO}+\frac{1}{e_r^4}\mathcal W^{\rm f-h\,NLO}\right.\nonumber\\
&&\left.
+\frac{\pi}{e_r^5}\mathcal W^{\rm f-h\,NNLO}+\frac{1}{e_r^6}\mathcal W^{\rm f-h\,N^3LO}
\right]\,,
\eea
with the various $\mathcal W^{\rm f-h\,nLO}$ listed in Table \ref{mathcal_W_f_min_h_table}.


\begin{table*}
\caption{\label{mathcal_W_f_min_h_table}  Coefficients $\mathcal W^{\rm f-h\,nLO}$ entering the large-$e_r$ expansion \eqref{Wfmenharer} of f-h contribution $W^{\rm f-h}$ to the first-order tail.
}
\begin{ruledtabular}
\begin{tabular}{|l|l|}
Coefficient & Expression   \\
\hline
$\mathcal W^{\rm f-h\,LO}$ &$\frac{1}{16} C_1+ \left(-\frac{7}{32}\nu C_1+\frac{7}{32} C_1+\frac{5}{128} D_1\right)\frac{\eta^2}{\bar a_r}$\\
$\mathcal W^{\rm f-h\,NLO}$&$ \frac45 C_1+\frac{4}{15} C_2+\left[\left(-\frac{2}{3} C_2-\frac{96}{35} C_1\right)\nu+\frac47 D_1+\frac25 C_2+\frac{4}{35} D_2+\frac{114}{35}C_1\right]\frac{\eta^2}{\bar a_r}$\\
$\mathcal W^{\rm f-h\,NNLO}$&$\frac{21}{32} C_1+\frac38 C_3+\frac38  C_2$\\
&$
+\left[\left(-\frac{141}{64} C_1-\frac{31}{32} C_2-\frac{9}{16} C_3\right)\nu+\frac{13}{16} C_2+\frac{1}{16} D_3+\frac{195}{64} C_1+\frac{3}{16} D_2-\frac{3}{16} C_3+\frac{135}{256} D_1\right]\frac{\eta^2}{\bar a_r}
$\\
$\mathcal W^{\rm f-h\,N^3LO}$&$\frac{64}{15}C_1+\frac{16}{3}C_3+\frac{16}{5}C_2$\\
&$
+ \left[\left(-\frac{128}{15} C_2-\frac{48}{5} C_3-\frac{1472}{105} C_1\right)\nu+\frac{16}{15} D_3+\frac{64}{35} D_2+\frac85  C_3+\frac{16}{15} D_4+\frac{136}{15} C_2+\frac{2336}{105} C_1+\frac{80}{21} D_1\right]\frac{\eta^2}{\bar a_r}
$\\
\hline
\end{tabular}
\end{ruledtabular}
\end{table*}

Using the {\it minimal} solution of the 5+6PN constraints
\bea
\label{CDmin}
C_1^{\rm min}&=&\frac{168}{5}\,,\qquad
 C_2^{\rm min}=0\,,\qquad
  C_3^{\rm min}=0
\,,\nonumber\\
D_1^{\rm min}&=&\frac{271066}{4725}+\frac{21736}{189}\nu
\,,\qquad
D_2^{\rm min}=-\frac{39712}{189}\nu
\,,\nonumber\\
D_3^{\rm min}&=&-\frac{68108}{945}\nu
\,,\qquad
D_4^{\rm min}=0\,,
\eea
the previous expression becomes
\begin{widetext}
\bea
W^{\rm f-h}_{\rm min}&=&\frac{M\nu^3}{\bar a_r^{9/2}}H_{\rm tot}\eta^2\left\{
\frac{\pi}{e_r^3}\left[
\frac{21}{10}+\left(-\frac{43207}{15120}\nu+\frac{580061}{60480}\right)\frac{\eta^2}{\bar a_r}
\right]
+\frac{1}{e_r^4}\left[
\frac{672}{25}+\left(-\frac{1668832}{33075}\nu+\frac{4703992}{33075}\right)\frac{\eta^2}{\bar a_r}
\right]\right.\nonumber\\
&&\left.
+\frac{\pi}{e_r^5}\left[
\frac{441}{20}+\left(-\frac{1732117}{30240}\nu+\frac{594173}{4480}\right)\frac{\eta^2}{\bar a_r}
\right]
+\frac{1}{e_r^6}\left[
\frac{3584}{25}+\left(-\frac{3267904}{6615}\nu+\frac{95857952}{99225}\right)\frac{\eta^2}{\bar a_r}
\right]
\right\}\,.
\eea

Re-expressing $\bar a_r$ and $e_r$ in terms of $\bar E$ and $j$ we find
\beq
\label{WfmenhEj}
W^{\rm f-h}=\frac{W^{\rm f-h}_3}{j^3}+\frac{W^{\rm f-h}_4}{j^4}
+\frac{W^{\rm f-h}_5}{j^5}+\frac{W^{\rm f-h}_6}{j^6}+O(j^{-7})\,,
\eeq
with the coefficients $W^{\rm f-h}_n$ listed in Table \ref{W_f_min_h_table} below.


\begin{table*}
\caption{\label{W_f_min_h_table}  Coefficients $W^{\rm f-h}_n$ entering the large-$e_r$ expansion \eqref{WfmenhEj} of f-h contribution $W^{\rm f-h}$ to the first-order tail.
}
\begin{ruledtabular}
\begin{tabular}{|l|l|}
$W^{\rm f-h}_n$ & Expression   \\
$W^{\rm f-h}_3$&$ \pi (2\bar E)^3 M^2\nu^3\eta^2\left[\frac{1}{16} C_1+\left( \frac{5}{64} C_1+\frac{5}{128} D_1  -\frac{15}{64}\nu C_1\right) (2\bar E)\eta^2 \right]
$\\
$W^{\rm f-h}_4$&$ (2\bar E)^{5/2}M^2\nu^3\eta^2\left\{\frac45 C_1+\frac{4}{15} C_2+\left[
\left(-\frac{241}{70} C_1-\frac{9}{10} C_2\right)\nu+\frac{4}{7} D_1+\frac{207}{70} C_1+\frac{4}{35} D_2+\frac{3}{10} C_2\right](2\bar E)\eta^2 \right\}
$\\
$W^{\rm f-h}_5$&$\pi (2\bar E)^2M^2\nu^3\eta^2\left\{
\frac{3}{8} C_2+\frac{9}{16} C_1+\frac{3}{8} C_3\right.$\\
&$\left.
+\left[\left(-\frac{9}{8} C_3-\frac{45}{16} C_1-\frac{49}{32} C_2\right)\nu+\frac{1}{16} D_3+\frac{3}{8} C_3+\frac{33}{8} C_1+\frac{11}{8} C_2+\frac{15}{32} D_1+\frac{3}{16} D_2\right](2\bar E)\eta^2\right\}$\\
$W^{\rm f-h}_6$&$ (2\bar E)^{3/2}M^2\nu^3\eta^2\left\{\frac{8}{3} C_1+\frac{8}{3} C_2+\frac{16}{3} C_3 \right.$\\
&$\left.
+\left[\left(-\frac{67}{5} C_2-\frac{314}{15} C_3-\frac{79}{5} C_1\right)\nu+\frac{103}{3} C_1+\frac{16}{15} D_3+\frac{8}{5} D_2+\frac{307}{15} C_2+\frac{98}{5} C_3+\frac{16}{15} D_4+\frac{8}{3} D_1\right](2\bar E)\eta^2 \right\}
$\\
\end{tabular}
\end{ruledtabular}
\end{table*}
Using the minimal value solutions of the $C_i$ and $D_i$ we find
\bea
W^{\rm f-h}_{\rm min}&=&\frac{(2\bar E)^3}{j^3}M^2\nu^3\eta^2\left\{
\pi\left[\frac{21}{10}+\left(\frac{294293}{60480}-\frac{10229}{3024}\nu\right)(2\bar E)\eta ^2\right]\right.\nonumber\\
&&
+\frac{(2\bar E)^{-1/2}}{j}\left[\frac{672}{25}+\left(\frac{4370596}{33075}-\frac{2446756}{33075} \nu \right)(2\bar E)\eta ^2\right]\nonumber\\
&&
+\frac{\pi(2\bar E)^{-1}}{j^2}\left[\frac{189}{10}+\left(\frac{834077}{5040}-\frac{22813}{270}\nu \right)(2\bar E)\eta ^2\right]\nonumber\\
&&\left.
+\frac{(2\bar E)^{-3/2}}{j^3}\left[\frac{448}{5}+\left(\frac{18520808}{14175}-\frac{143384}{225}\nu\right)(2\bar E)\eta ^2\right]
\right\}\,.
\eea

\subsection{Second-order-tail part}

Finally, the second-order-tail contribution turns out to be
\beq
W^{\rm tail, h, 5.5PN}= \frac{M^2\nu^2}{e_r^4\bar a_r^5}\left[\frac{23968}{675}+\frac{10593}{1400}\frac{\pi^3}{e_r}+\left(\frac{835456}{4725}+\frac{4738816}{70875}\pi^2\right)\frac1{e_r^2}\right]\,,
\eeq
or equivalently
\beq
W^{\rm tail, h, 5.5PN}= \frac{M^2\nu^2(2\bar E)^2}{j^4}\left[\frac{23968}{675}(2\bar E)+\frac{10593}{1400}\frac{\pi^3(2\bar E)^{1/2}}{j}+\left(\frac{499904}{4725}+\frac{4738816}{70875}\pi^2\right)\frac{1}{j^2}\right]\,.
\eeq

\end{widetext}


\end{document}